\RequirePackage[loading]{tracefnt}
\documentclass[a4paper,fleqn,usenatbib,useAMS]{mnras}

\usepackage[T1]{fontenc}
\usepackage{ae,aecompl}

\usepackage{epsfig}
\usepackage{amsmath}
\usepackage{times}
\usepackage{graphicx}
\usepackage{lineno}
\usepackage{float}
\usepackage{url}
\usepackage{amssymb}
\usepackage{enumitem}
\usepackage{subcaption}
\usepackage[flushleft]{threeparttable}
\usepackage{enumitem}
\setbox1=\hbox{$\bullet$}
\setlist{noitemsep, topsep=0pt, parsep=0pt, partopsep=0pt, leftmargin=0.7cm, labelindent=0.3cm, labelwidth=\wd1, itemindent=*, labelsep=\dimexpr0.3cm-\wd1}

\title[A low-frequency transient near the NCP] 
{LOFAR MSSS: Detection of a low-frequency radio transient in 400~hrs of monitoring of the North Celestial Pole}

\author[Stewart et al.]
       {A. J. Stewart,$^{1,2}$\thanks{email:adam.stewart@physics.ox.ac.uk} R. P. Fender,$^{1,2}$ J. W. Broderick,$^{1,2,3}$ T. E. Hassall,$^{1,2}$ T. Mu\~{n}oz-Darias,$^{4,5,1,2}$
\newauthor A. Rowlinson,$^{6,3}$ J. D. Swinbank,$^{7,6}$ T. D. Staley,$^{1,2}$ G. J. Molenaar,$^{6,8}$ B. Scheers,$^{9,6}$
% ...
\\
	\\Affiliations are listed at the end of the paper}
\date{Accepted 2015 November 25.  Received 2015 November 24; in original form 2015 July 17}
\pubyear{2015}

\begin{document}
\label{firstpage}
\pagerange{\pageref{firstpage}--\pageref{lastpage}}
\maketitle

\begin{abstract}
We present the results of a four-month campaign searching for low-frequency radio transients near the North Celestial Pole with the Low-Frequency Array (LOFAR), as part of the Multifrequency Snapshot Sky Survey (MSSS). The data were recorded between 2011 December and 2012 April and comprised 2149 11-minute snapshots, each covering 175 deg$^{2}$. We have found one convincing candidate astrophysical transient, with a duration of a few minutes and a flux density at 60 MHz of 15--25 Jy. The transient does not repeat and has no obvious optical or high-energy counterpart, as a result of which its nature is unclear. The detection of this event implies a transient rate at 60 MHz of $3.9^{+14.7}_{-3.7}\times10^{-4}$ day$^{-1}$ deg$^{-2}$, and a transient surface density of $1.5\times10^{-5}$ deg$^{-2}$, at a 7.9-Jy limiting flux density and $\sim10$-minute time-scale. The campaign data were also searched for transients at a range of other time-scales, from 0.5 to 297 min, which allowed us to place a range of limits on transient rates at 60 MHz as a function of observation duration.
\end{abstract}




\section{Introduction}
\label{sec:intro}
The variable and transient sky offers a window into the most extreme events that take place in the Universe. Transient phenomena are observed at all wavelengths across a diverse range of objects, ranging from optical flashes detected in the atmosphere of Jupiter caused by bolides \citep{jupiter}, to violent Gamma-Ray Bursts (GRBs) at cosmological distances which can outshine their host galaxy \citep{grb1,grb2}. Observations at radio wavelengths provide a robust method to probe these events, supplying unique views of kinetic feedback and propagation effects in the interstellar medium, which are also just as diverse in their associated time-scales. Active Galactic Nuclei \citep[AGN;][]{agn1,agn2} are known to vary over time-scales of a month or longer, whereas observations of the Crab Pulsar have seen radio bursts with a duration of nanoseconds \citep{crab}.\\

Historically, and still to this day, radio observations have been used to follow-up transient detections made at other wavelengths. Radio facilities generally had a narrow field-of-view (FoV), which made them inadequate to perform rapid transient and variability studies over a large fraction of the sky. However, blind transient surveys have been performed and have produced intriguing results. For example, \citet{bower} \citep[also see][]{frail} discovered a single epoch millijansky transient at 4.9 GHz while searching 944 epochs of archival Very Large Array (VLA) data spanning 22 years, with three other possible marginal events. Sky surveys using the Nasu Observatory have also been successful in finding a radio transient source, with \citet{Nasu2} having observed a two epoch event, peaking at 3 Jy at 1.42 GHz. Various counterparts were considered at other wavelengths, but the origin of the transient remains unknown. Lastly, \citet{bannister} surveyed 2775 deg$^2$ of sky at 843 MHz using the Molonglo Observatory Synthesis Telescope (MOST), yielding 15 transients at a 5$\sigma$ level of 14 mJy beam$^{-1}$, 12 of which had not been previously identified as transient or variable.\\

Surveys at low frequencies ($\leq$ 330 MHz) have also been completed. \citet{lazio} carried out an all-sky transient survey using the Long Wavelength Demonstrator Array (LWDA) at 73.8 MHz, which detected no transient events to a flux density limit of 500 Jy. In addition, \citet{Hym02,Hym05,Hym06,Hym09} discovered three radio transients during monitoring of the Galactic centre at 235 and 330 MHz. These were identified by using archival VLA observations along with regular monitoring using the VLA and the Giant Metrewave Radio Telescope (GMRT). The transients had flux densities in the range of 100 mJy--1 Jy and occurred on time-scales ranging from minutes to months. Lastly, \citet{jaeger} searched six archival epochs from the VLA at 325 MHz centred on the \textit{Spitzer-Space-Telescope} Wide-Area Infrared Extragalactic Survey (SWIRE) Deep Field. In an area of 6.5 deg$^2$ to a 10$\sigma$ flux limit of 2.1 mJy beam$^{-1}$, one day-scale transient event was reported with a peak flux density of 1.7 mJy~beam$^{-1}$.\\

Radio transient surveys are being revolutionised by the development of the current generation of radio facilities. These include new low-frequency instruments such as the International Low-Frequency Array \citep[LOFAR;][]{lofar}, Long Wavelength Array \citep[LWA;][]{lwa} and the Murchinson Wide Field Array \citep[MWA;][]{mwa}. The telescopes listed offer a large FoV coupled with an enhanced sensitivity, with LOFAR having the capability to reach sub-mJy sensitivities and arcsecond resolutions (though this full capability is not used in this work as such modes were being commissioned at the time). These features are achieved by utilising phased-array technology with omnidirectional dipoles, and the before mentioned telescopes act as pathfinders for the low-frequency component of the Square Kilometre Array \citep[SKA;][]{ska}. With such greatly improved sensitivities at low frequencies, we have a new opportunity to survey wide areas of the sky for transients and variables, with a particular sensitivity to coherent bursts.

These new facilities have already produced some interesting results in this largely unexplored parameter space. \citet{Bell} searched an area of 1430 deg$^2$ for transient and variable sources at 154 MHz using the MWA. No transients were found with flux densities $>$ 5.5 Jy on time-scales of 26 minutes and one year. However, two sources displayed potential intrinsic variability on a one year time-scale. Using the LWA, \citet{lwatrans1} detected two kilojansky transient events while using an all-sky monitor to search for prompt low-frequency emission from GRBs. They were found at 37.9 and 29.9 MHz, lasting for 75 and 100 seconds respectively, and were not associated with any known GRBs. This was followed up by \citet{lwatrans2} who searched over 11\,000 hours of all-sky images for similar events, yielding 49 candidates, all with a duration of tens of seconds. It was discovered that 10 of these events correlated both spatially and temporally with large meteors (or fireballs). This low-frequency emission from fireballs was previously undetected and identifies a new form of naturally occurring radio transient foreground.

Two transient studies have now also been completed using LOFAR. \citet{Dario} searched 2275 deg$^2$ of sky at 150 MHz, at cadences of 15 minutes and several months, with no transients reported to a flux limit of 0.5 Jy. \citet{Cendes} searched through 26, 149-MHz observations centred on the source Swift J1644+57, covering 11.35 deg$^2$. No transients were found to a flux limit of 0.5 Jy on a time-scale of 11 minutes.\\

In this paper we use the LOFAR telescope to search 400 hours of observations centred at the North Celestial Pole (NCP; $\delta=$ 90$^\circ$), covering 175 deg$^2$ with a bandwidth of 195 kHz at 60 MHz. LOFAR is a low-frequency interferometer operating in the frequency ranges of 10--90 MHz and 110--250 MHz. It consists of 46 stations: 38 in the Netherlands and 8 in other European countries. Full details of the instrument can be found in \citet{lofar}.

\begin{figure*}
\includegraphics[scale=0.35]{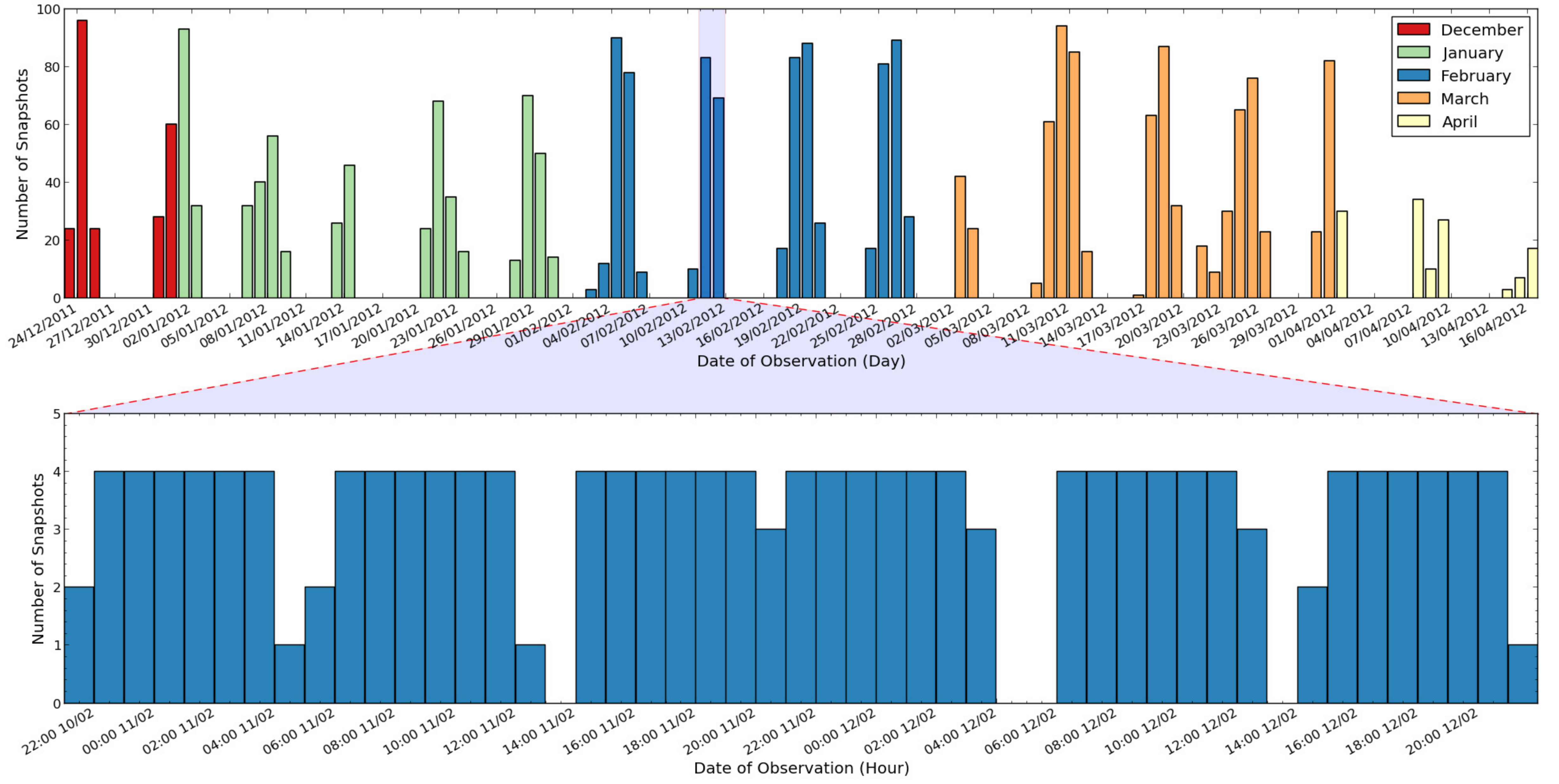}
\caption{Histograms giving a general overview of when the 2609 11-minute snapshots of the NCP were observed. The top panel contains a histogram showing how many snapshots were observed on each day over the entire four month period, colour coded by month, which shows the distinct observing blocks in which NCP observations were obtained. The bottom panel displays a `zoom-in' of the date range 21:00 2012/02/10--21:00 2012/02/12 \textsc{utc}, now showing the number of snapshots per hour. This emphasises further the sometimes fragmented nature of the observing pattern of the NCP, with which careful consideration had to be given on how to combine the observations for the transient search.} 
\label{fig:dates}
\end{figure*}

A previous study of variable radio sources located near the NCP field ($75^\circ < \delta < 88^\circ $) was carried out by \citet{russncp}. This study identified 15 objects displaying variability at centimetre wavelengths on time-scales of days or longer. However, the variability amplitude was found to be within seven per cent, which we would not be able to distinguish with LOFAR due to general calibration uncertainties at the time of writing. In addition, the lower observing frequency used in this work would mean that the expected peak flux densities would be significantly lower, assuming a standard synchrotron event \citep[e.g.,][]{vdl}, making them challenging to detect. Also, the lower frequency means that the variability would occur over even longer time-scales, again assuming that the emission arises from a synchrotron process.\\

The observations and processing techniques are discussed in Section~\ref{sec:obs}, with a description of how the transient search was performed in Section~\ref{sec:trap}. The results can be found in Section~\ref{sec:results}, which is followed by a discussion of a discovered transient event in Section~\ref{sec:transient1}. The implied transient rates and limits are discussed in Section~\ref{sec:limits}, before we conclude in Section~\ref{sec:conclusion}.

\section{LOFAR Observations of the NCP}
\label{sec:obs}
The monitoring survey of the NCP was performed between 2011 December 23--2012 April 16, resulting in a total of 2609 observations being recorded. The NCP was chosen because it is constantly observable from the Northern Hemisphere, and the centre of the field is located towards constant azimuth and elevation (az/el) coordinates. However, this is not true for sources which lie away from the NCP, where these sources rotate within the LOFAR elliptical beam. We therefore restrict our transient search to an area around the NCP where the LOFAR station beam properties are consistent for each epoch observed, avoiding systematic errors in the light curves that might be introduced if this was not the case. It is also an advantage that the line-of-sight (\textit{b}=122$^{\circ}$.93, \textit{l}=+27$^{\circ}$.13) is located towards a relatively low column density of Galactic free electrons; the maximum expected dispersion measure (DM) is 55\,pc\,cm$^{\rm{-3}}$ according to the NE2001 model of the Galactic free electron distribution \citep{dm}.

The NCP measurements were taken using the LOFAR Low-Band Antennas (LBA) at a single frequency of 60 MHz; the bandwidth was 195 kHz, consisting of 64 channels. The total integration time of each snapshot was 11 min, sampled at 1 s intervals, and data were recorded using the `LBA\_INNER' setup, where the beam is formed using the innermost 46 LBA antennas from each station, which gives the largest possible FoV and a full width half maximum (FWHM) of 9.77$^{\circ}$.

\subsection{Observation epochs}
\label{sec:epochs}
The programme piggybacked on another commissioning project being performed by LOFAR at the time, the Multifrequency Snapshot Sky Survey (MSSS) -- the first major LOFAR observing project surveying the low-frequency sky \citep{MSSS}. With every single MSSS LBA observation that took place, a beam was placed on the NCP using one subband of the full observational setup for MSSS. Figure~\ref{fig:dates} shows a histogram of the number of NCP snapshots observed each day over the duration of the programme, in addition to a similar histogram showing the number of snapshots per hour for a particular set of days. Of the 2609 snapshots, 909 were recorded during the day and 1700 were recorded at night. The MSSS observational set-up also meant that each 11-minute snapshot in the same observation block was separated by a time gap of four minutes. 

\subsection{Calibration and imaging}
\label{sec:imaging}
Prior to any processing, radio-frequency interference (RFI) was removed using {\sc aoflagger} \citep{aoflag1,aoflag2,aoflag3} with a default strategy, in addition, the two channels at the highest, and lowest, frequency edges of the measurement set were also completely flagged, reducing the bandwidth to 183 kHz. When using an automatic flagging tool such as {\sc aoflagger}, it is important to be aware of the fact that transient sources could be mistakenly identified as RFI by the software. This is a complex issue which is beyond the scope of this work. However, an initial investigation for the LOFAR case was carried out by \citet{Cendes}. In these tests, simulated transient sources, described by a step function, with different flux densities and time durations (from seconds to minutes), were injected into an 11 minute dataset. These datasets were subsequently passed through {\sc aoflagger} before calibrating and imaging as normal in order to observe how the simulated transient was affected by the automatic flagging, if at all. The authors concluded that transient signals shorter than a duration of two minutes could be partially, or in the case of $\sim$Jansky level sources, completely flagged. However, there are some caveats to this testing: short time-scale imaging was not tested for short-duration transients, and it remains to be determined how the automatic flagging would treat other types of transients (i.e. a non step function event). Hence, while these results certainly suggest that transients could be affected by {\sc aoflagger}, further testing is required to completely understand how automatic flagging software can affect the detection of a transient.

At this stage we also removed all data from international LOFAR stations, leaving just the Dutch stations. This was due to the complex challenges in reducing these corresponding data at the time of processing. Following this, the `demixing' technique \citep[described by][]{demix} was used to remove the effects of the bright sources Cassiopeia A and Cygnus A from the visibilities. Finally, averaging in frequency and time was performed such that each observation consisted of 1 channel and an integration time of 10 seconds per time step. The averaging of the data was necessary to reduce the data volume and computing time required to process the data.

This averaging has the potential to introduce effects caused by bandwidth and time smearing, which are discussed in more detail by \citet{MSSS} in relation to MSSS data. Following \citet{MSSS}, we used the approximations given by \citet{smearing} to calculate the magnitude of the flux loss $(S/S_0)$ in each case, assuming a projected baseline length of 10 km. We found the bandwidth smearing factor to equal seven per cent (using a field radius corresponding to the FWHM) and a time smearing factor of 0.4 per cent. Thus, while the effect of time smearing was negligible, the impact of bandwidth smearing was potentially significant, yet remained within the calibration error margins (10 per cent; see Section~\ref{sec:imagequality}).\\
\\
A selection of flux calibrators, characterised by \citet{HealdCalibs}\footnote{Cygnus A is not characterised by \citet{HealdCalibs}, but extensive commissioning work (summarised by \citealt{CygA} and McKean et al. in prep.) has produced a detailed source model.}, were used in the main processing of the data and were observed simultaneously utilising LOFAR's multi-beam capability (thus the calibrator scans were also 11 minutes in length). The calibrators and their usage can be found in Table~\ref{table:calibs}. The standard LOFAR imaging pipeline was then implemented which consists of the following steps. Firstly, the amplitude and phase gain solutions, using XX and YY correlations, are obtained for each calibrator observation using Black Board Selfcal \citep[{\sc bbs};][]{BBS}. These solutions are direction-independent, and are derived for each time step using the full set of visibilities from the Dutch stations, as well as a point source model of the calibrator itself. Beam calibration was also enabled which accounts, and corrects, for elevation and azimuthal effects with the station beam. The amplitudes of these gain solutions were then clipped to a $3\sigma$ level to remove significant outliers, which were not uncommon in these early LOFAR data. The gain solutions were then transferred directly from the calibrators to the respective NCP observation. 

Secondly, a phase-only calibration step was performed (also using {\sc bbs}) to calibrate the phase in the direction of the target field. The solutions were derived using data within a maximum projected \textit{uv} distance of $4000\lambda$ (20 km; 24 core + 10 remote stations). In order to perform this step, a sky model was obtained of the NCP field using data from the global sky model (GSM) developed by \citet{scheers}. This model is constructed by firstly gathering sources which are present within a set radius from the target pointing in the 74 MHz VLA Low-Frequency Sky Survey (VLSS; \citealt{VLSS}). In the NCP case, the radius was set to 10 deg. From this basis, sources are then cross-correlated, using a source association radius of 10 arcsec, with the 325 MHz Westerbork Northern Sky Survey \citep[WENSS;][]{WENSS} and the 1400 MHz NRAO VLA Sky Survey \citep[NVSS;][]{NVSS} to obtain spectral index information. In those cases where no match was found, the spectral index, $\alpha$ (using the definition $S_\nu\propto\nu^\alpha$), was set to a canonical value of $\alpha = -0.7$. No self-calibration was performed on the data. The reader is referred to \citet{lofar} for more LOFAR standard pipeline information.

The main MSSS project discovered that observations recorded during this 2011-2012 period potentially contained one or more bad stations, and the data quality would improve if such stations were removed. LOFAR was still very much in its infancy at the time, and, as a result, was not entirely stable; problems such as network connection issues or bad digital beam forming contributed to the poor performance of some stations. Hence, an automated tool was developed which analysed each station, identifying and flagging those that displayed a significant number of baselines with high measured noise. This tool was utilised in the NCP processing and primarily removed stations with poorly-focussed beam responses \citep{MSSS}. It should be noted that present LOFAR data no longer require this tool as the issues outlined above have been rectified.

Finally, a FoV of 175 deg$^{2}$ was imaged using the {\sc awimager} \citep{AW}, with a robust weighting parameter of 0 \citep{briggs}, and a primary-beam (PB) correction applied to each image. A maximum projected baseline length of 10 km was used in this study ($2000\lambda$; 24 core + 7 remote stations). This was chosen to obtain good \textit{uv} coverage and a maximum resolution for which we were confident with the calibration. The typical resolution for the 11-minute snapshots was 5.4 $\times$ 2.3 arcmin.

\begin{table}
\centering
\caption{Table listing the calibrators used for the NCP observations. It was decided early in the MSSS programme that 3C 48 and 3C 147 might not be adequate as calibrators for the LBA portion of the survey, and so these were dropped 8 and 22 days after first use, respectively. Observations using these calibrators displayed no disadvantages over those observed with other calibrators when checked in this project, and hence they were kept as part of the sample.}
\label{table:calibs}
\begin{tabular}{|c|c|c|c|}
\hline
Calibrator Source & \% Use & First Use Date & Last Use Date\\
\hline
\hline
3C 48 & 2\% & 2011 Dec 24 & 2012 Jan 01\\
3C 147 & 6\% & 2011 Dec 23 & 2012 Jan 14\\
3C 196 & 43\% & 2011 Dec 24 & 2012 Apr 14\\
3C 295 & 40\% & 2011 Dec 24 & 2012 Apr 01\\
Cygnus A & 9\% & 2012 Jan 28 & 2012 Apr 16\\
\hline
\end{tabular}
\end{table}

\subsection{Quality control}
\label{sec:quality}
A number of bad-quality observations were detected and subsequently flagged using two methods: (i) checking the processed visibilities and (ii) inspecting the final images for each 11-minute snapshot. When analysing the visibilities, poor snapshots were flagged when the calibrated visibilities had a mean value greater than the overall mean of the entire four month dataset plus one standard deviation value.  A slight, or indeed dramatic rise in the mean of the visibilities does not necessarily imply a completely bad dataset: an extremely bright transient ( $>$ 100 Jy) could have this effect, for example. Such events may have been previously seen from flare stars at low frequencies \citep{abdul}, although at shorter time-scales than 11 minutes ($\sim$1 s). However, overall, the survey is less sensitive to extremely bright events because of this quality control step. It was beyond the scope of this project to fully investigate this possible effect, and so we decided to only use measurement sets that were deemed to be sufficiently well calibrated.

The results from the automated flagging were also checked against a manual analysis of the visibility plots and the snapshot images, the latter enabling the detection of more bad observations. In total, 460 (out of 2609) snapshots were marked as bad, and were discarded from the search. The large size of the full dataset meant that there was no single common reason as to why individual snapshots were rejected, but the problems that caused rejection were mostly due to RFI or ionospheric issues. After the quality control was completed, 2149 observations (394 hr) were considered in the analysis.

\section{Transient \& Variability Search Method}
\label{sec:trap}

\subsection{Time-scales searched}
\label{sec:timescales}
As the properties of the target transient population are unknown, the complete dataset was split and combined in various ways to fully explore the transient parameter space available. Along with performing a search on the original snapshots, each with an integration time of 11 minutes, searches were also performed on images with integration times of 30 seconds, 2 minutes, 55 minutes and 297 minutes. For the longer-duration images, only those 11-minute snapshots which were four minutes apart were combined together and imaged. This was to keep the visibilities as continuous as possible in the search for transients. After the quality control step described in Section~\ref{sec:quality}, 297 minutes was the longest continuous integration time possible. All calibration was performed on each individual 11-minute snapshot; for the longer time-scales the relevant datasets were combined and then imaged.

\begin{figure*}
\centering
{\includegraphics[scale=0.53]{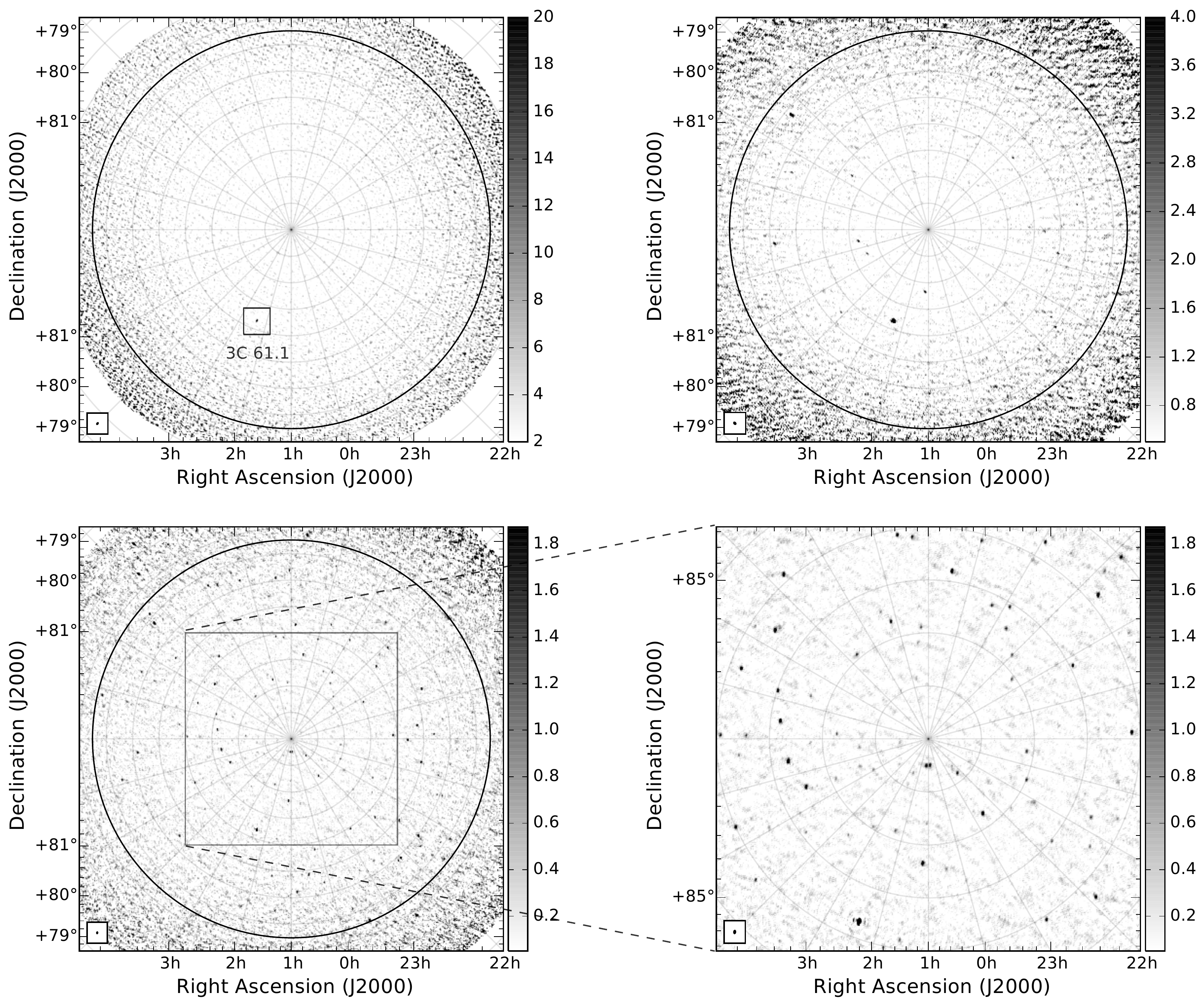}}
\caption{Examples of the NCP field maps at different time-scales. Where present, the area within the black circle indicates the portion of the image searched for transients. This was the same for each time-scale and had a radius of $7.5^{\circ}$. \textit{Upper left panel}: an image on the 30 s time-scale which was observed on 2012 January 9. Using projected baselines of up to 10 km, the map has a resolution of 4.2 $\times$ 2.3 arcmin (synthesized beam position angle [BPA] $-39^\circ$) with a noise level of 1.9 Jy beam$^{-1}$. Only the source 3C\,61.1 is detected at a 10$\sigma$ level, and this source is marked on the image. \textit{Upper right panel}: an 11 minute snapshot observed on 2011 December 31. The noise level is 320 mJy beam$^{-1}$ and the resolution is 5.6 $\times$ 3.6 arcmin (BPA $43^\circ$). The number of detected sources at a 10$\sigma$ level is now $\sim15$. \textit{Lower left panel}: an example of the longest time-scale images available of 297 minutes, constructed by concatenating and imaging 27, 11-minute sequential snapshots. Observed on 2012 February 4, this image has a resolution of 3.5 $\times$ 2.0 arcmin (BPA $-6^\circ$) and a noise level of 140 mJy beam$^{-1}$, with $\sim50$ sources now detected at a 10$\sigma$ level. \textit{Lower right panel}: a magnified portion of the lower left panel image. The colour bar units are Jy beam$^{-1}$.}\label{fig:ncpall}
\end{figure*}

\subsection{The Transients Pipeline}
\label{sec:thetrap}
The analysis of the data and search for radio transients was performed using software developed by the LOFAR Transients Key Science Project, named the Transients Pipeline ({\sc trap}). It is built to search for transients in the image plane, whilst also storing light curves and variability statistics of all detected sources. Moreover, it is designed to cope with large datasets containing thousands of sources such as this NCP project. A full and detailed overview of the {\sc trap} can be found in \citet{TraP}\footnote{The work presented in this paper primarily used {\sc trap} release 1.0. However, the data were re-processed once {\sc trap} release 2.0 was available, which is the version described by \citet{TraP}, to confirm results.}. In brief it performs the following steps:
\begin{enumerate}[leftmargin=*]
\item Input images are passed through the {\sc trap} quality control which examines two features of the images. Firstly, the rms of the map is compared against the expected theoretical rms of the observation, and if the ratio between the observed and theoretical rms is above a set threshold then the image is flagged as bad. In this case, the threshold was set to the mean ratio value of each time-scale plus one standard deviation. The second test involves checking that the beam is not excessively elliptical by comparing the ratio of the major and minor axes. If this value is over a set threshold then the image is also flagged as bad. All bad images are then rejected and are not analysed by the {\sc trap} (see Rowlinson et al. in prep. for methods of setting these thresholds). The number of images accepted by the {\sc trap} compared to the total entered can be seen in Table~\ref{table:times}.
\item Sources are extracted using {\sc pyse} - a specially developed source extractor for use in the {\sc trap} \citep[][Carbone et al. in prep]{pyse}. Importantly, all sources are initially extracted as unresolved point sources, which would be expected from a transient event.
\item For each image, the source extraction data are analysed to associate each source with previous detections of the same source, such that a light curve is constructed. In cases where no previous source is associated with an extraction, the source is flagged as a potential `new source' and is continually monitored from the detection epoch onwards.
\end{enumerate}
For the source extraction, we define an island threshold, which defines the region in which source fitting is performed, and a detection threshold where only islands with peaks above this value are considered. These island and detection thresholds were set to 5$\sigma$ and 10$\sigma$ respectively. While the use of a 10$\sigma$ detection threshold may seem very conservative, we agree with the arguments presented by \citet{10sigma} (hereafter MWB15) who advocate this criteria when identifying a transient source. In their paper, the authors' main motivation for this high threshold is the significant possibility of spurious signals such as those seen in previous radio transient searches \citep{GalYam,Ofek2010,Croft,frail,Nasu7}, arising from calibration artefacts, residual sidelobes and other similar issues. We share these concerns, in addition to be being generally cautious as this survey is one of the first conducted with the new LOFAR telescope. As also stated by MWB15, previous surveys have used $5\sigma$ as a detection threshold, which will of course increase the number of potential transient detections; however, this will also yield a high number of false detections, especially with the large number of epochs being used in this survey. Thus, minimising false detections and obtaining a manageable number of transient candidates were further motivations to use a $10\sigma$ detection threshold. We refer the reader to MWB15 for further discussion on this topic.

The transient search was also constrained to within a circular area of radius 7.5 deg from the centre of the image. This was to avoid the outer part of the image which was much noisier and did not have reliable flux calibration.\\

For each lightcurve, two values are calculated in order to define whether a source is a likely transient or variable: $V_\nu$, a coefficient of variation, and $\eta_\nu$, the significance of the variability \citep{scheers}. $V_\nu$ is defined as
\begin{equation}	
\label{eq:variability}
V_\nu = \frac{s_\nu}{\overline{I_\nu}}=\frac{1}{\overline{I_\nu}}\sqrt{\frac{N}{N-1}\left(\overline{I^2_\nu}-{\overline{I_\nu}}^2\right)} \textrm{ ,}
\end{equation}
where $s$ is the unbiased sample flux standard deviation, $\overline{I}$ is the arithmetic mean flux of the sample, and $N$ is the number of flux measurements obtained for a source. The significance value, $\eta_\nu$, is based on reduced $\chi^2$ statistics and indicates how well a source lightcurve is modelled by a constant value. It is given by
\begin{equation}
\label{eq:secondvariability}
\eta_\nu = \frac{N}{N-1}\left(\overline{\omega I^{2}_\nu}-\frac{{\overline{\omega I_\nu}}^2}{\overline{\omega}}\right) \textrm{ ,}
\end{equation}
where $\omega$ is a weight which is inversely proportional to the error of a given flux measurement ($\omega=1/\sigma^2_{I_\nu}$). Throughout this paper we define these parameters as the `variability parameters'. For more detailed discussion on these parameters we refer the reader to \citet{scheers} and \citet{TraP}.

To define a transient or variable source, a histogram of each parameter for the sample was created and fitted with a Gaussian in logarithmic space. Any source which exceeds a 3$\sigma$ threshold on these plots is flagged as a potential candidate. Rowlinson et al. (in prep.) will offer an in-depth discussion on finding transient and variable sources using these methods.

\section{Results}
\label{sec:results}
\begin{figure*}
\centering
{\includegraphics[scale=0.46]{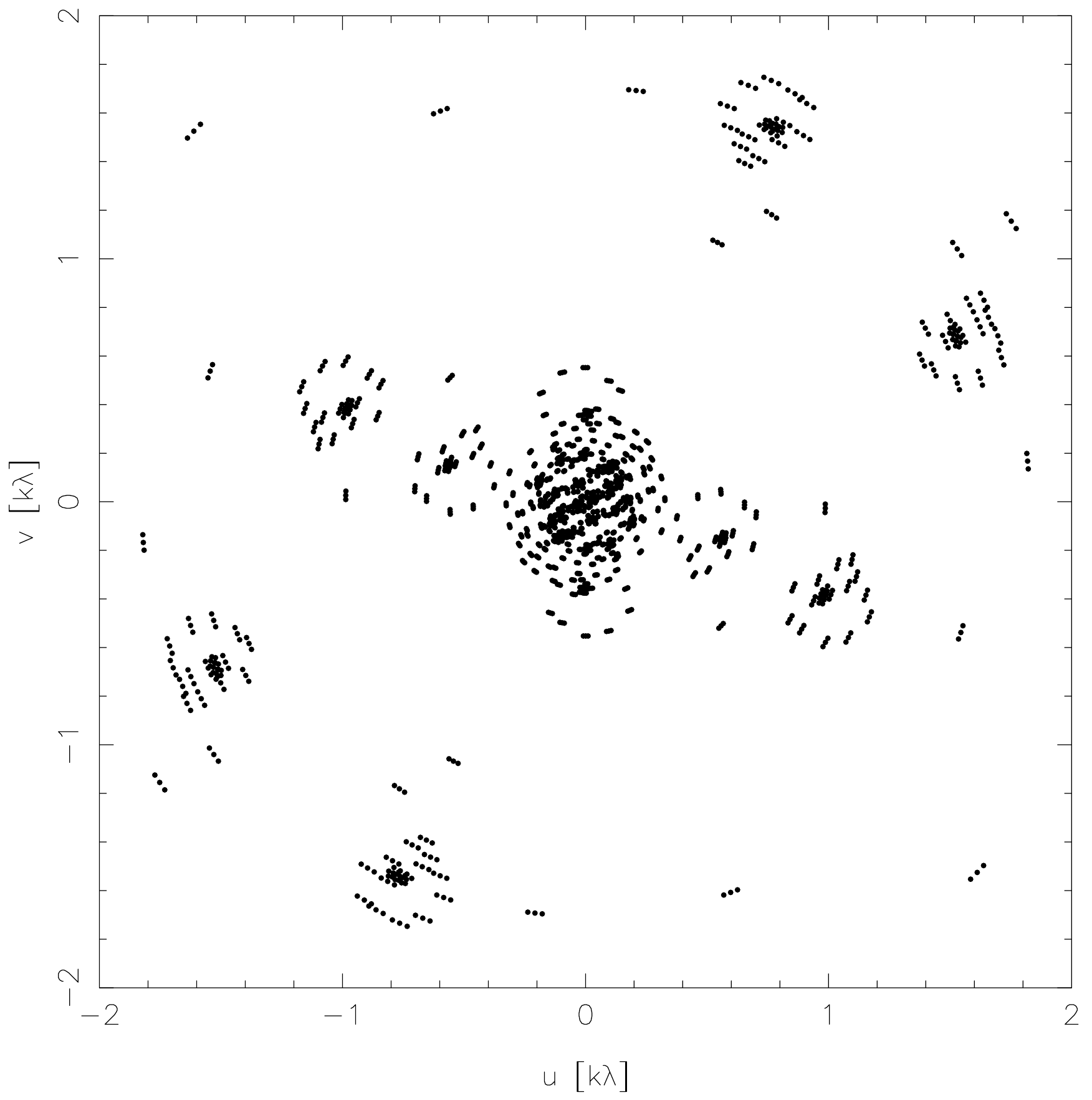}}
\hspace{0.5cm}
{\includegraphics[scale=0.46]{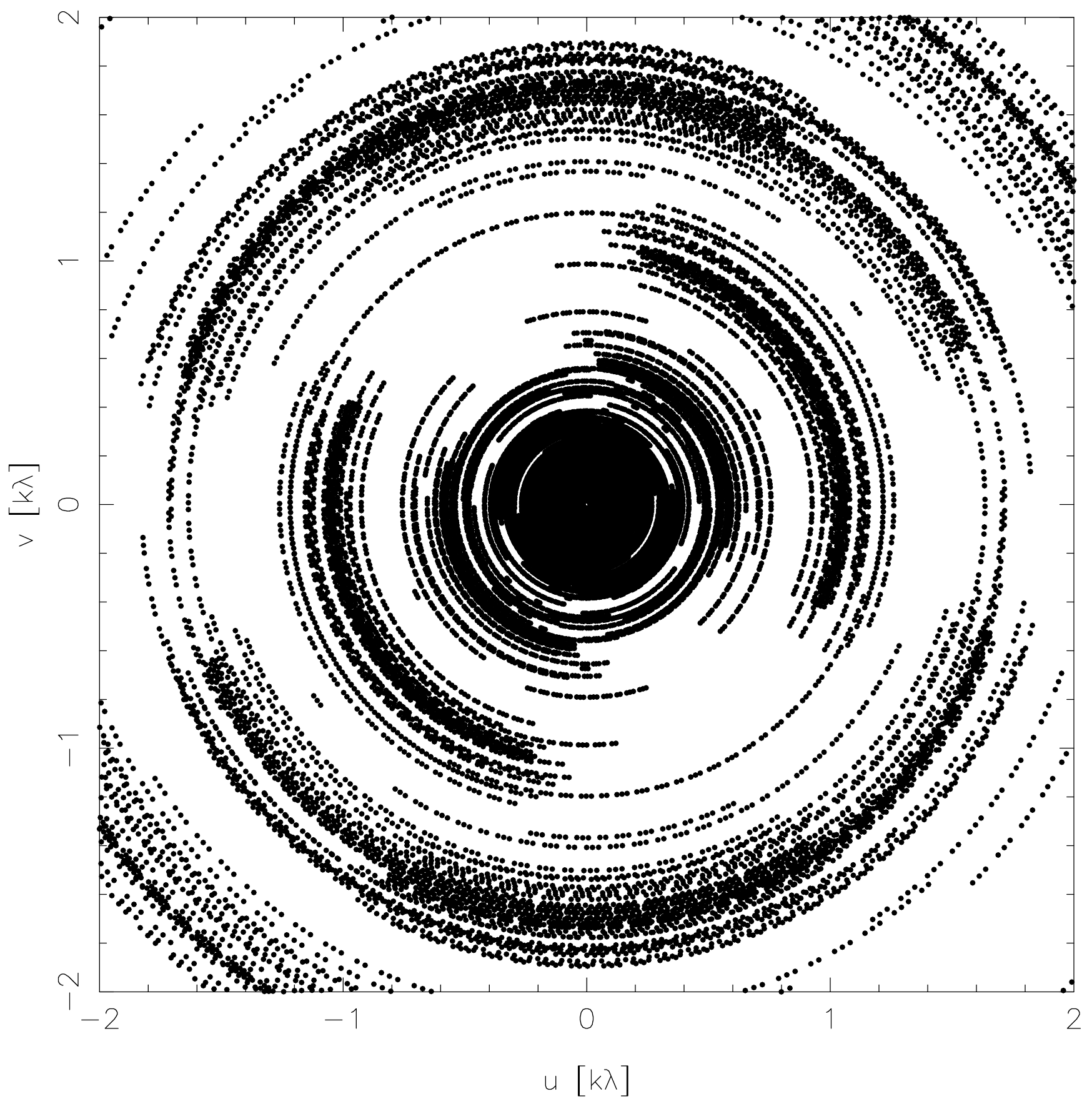}}
\caption{\textit{Left panel}: The \textit{uv} coverage obtained with an 11 min snapshot. \textit{Right panel}: The improved \textit{uv} coverage gained when combining 27 snapshots (297 min). In each case the \textit{uv} range is limited to $\pm$2 k$\lambda$ (10 km).}\label{fig:ncpuv}
\end{figure*}

\begin{figure}
\centering
{\includegraphics[width=\columnwidth]{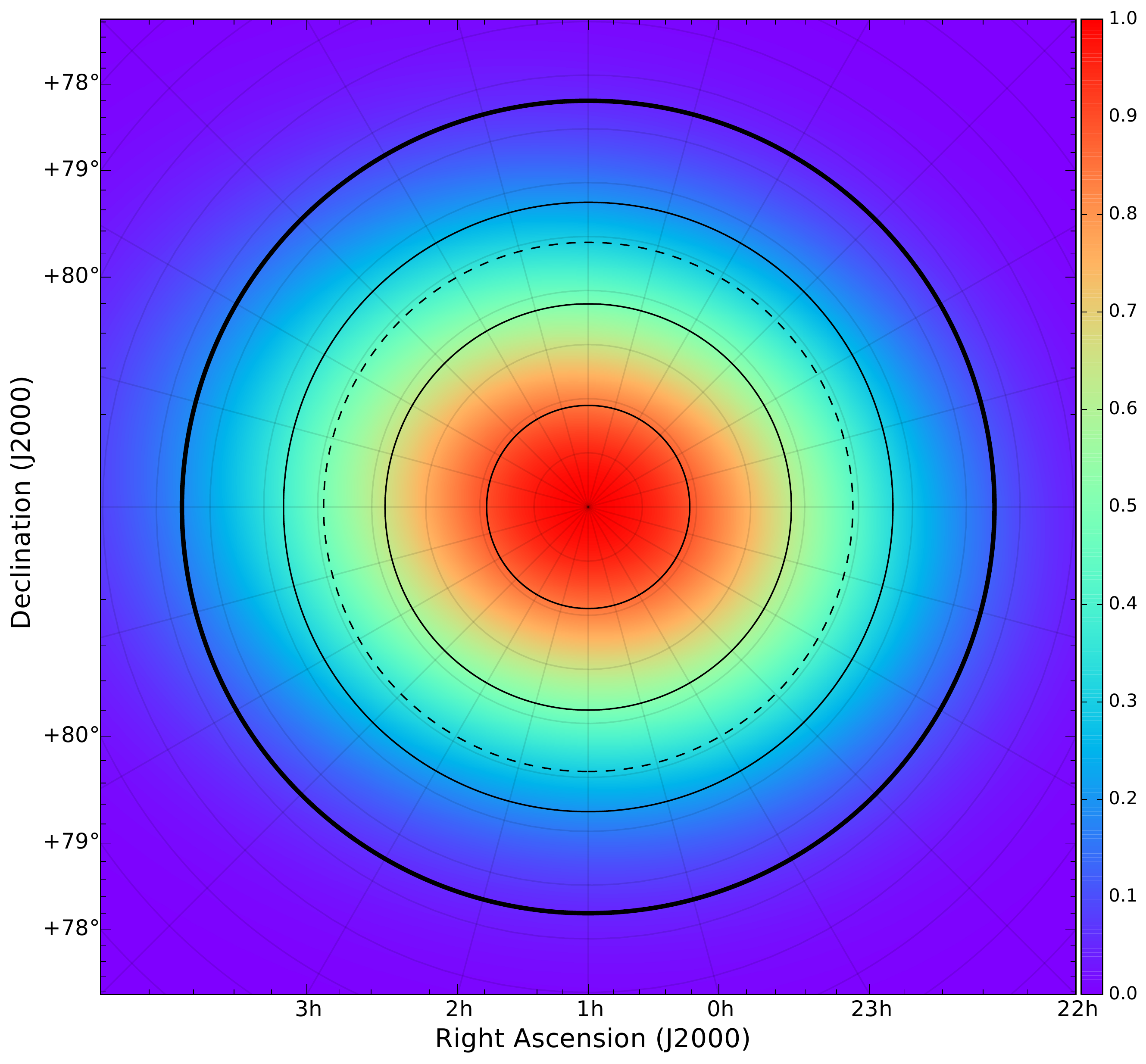}}
\caption{An example of a normalized primary beam map from one of the NCP observations, which has been scaled to 1.0. The bold, outer solid-line circle represents the full extent of the area for which the transient search was performed (radius of 7.5 deg). The inner solid-line circles show how the area was divided in order to gain an estimate of the average rms for each image accounting for the primary beam. The dashed-line circle indicates the position of the primary beam half-power point.}\label{fig:ncpPB}
\end{figure}
\subsection{Image quality}
\label{sec:imagequality}
Examples of the 30 s, 11 min and 297 min time-scale images can be found in Figure~\ref{fig:ncpall}. Note that imaging the NCP can sometimes cause confusion when displaying the right ascension (RA) and declination (Dec) on the image axis, as the grid lines become circular. The grid lines are shown in all figures to help demonstrate this. The obtained \textit{uv} coverage of the 11 and 297 min observations can be viewed in Figure~\ref{fig:ncpuv}. The average sensitivity reached with each time-scale is summarised in Table~\ref{table:times}, along with the number of epochs available after the quality control described in Sections~\ref{sec:obs} and~\ref{sec:trap}.

\begin{figure}
\centering
\includegraphics[scale=0.47]{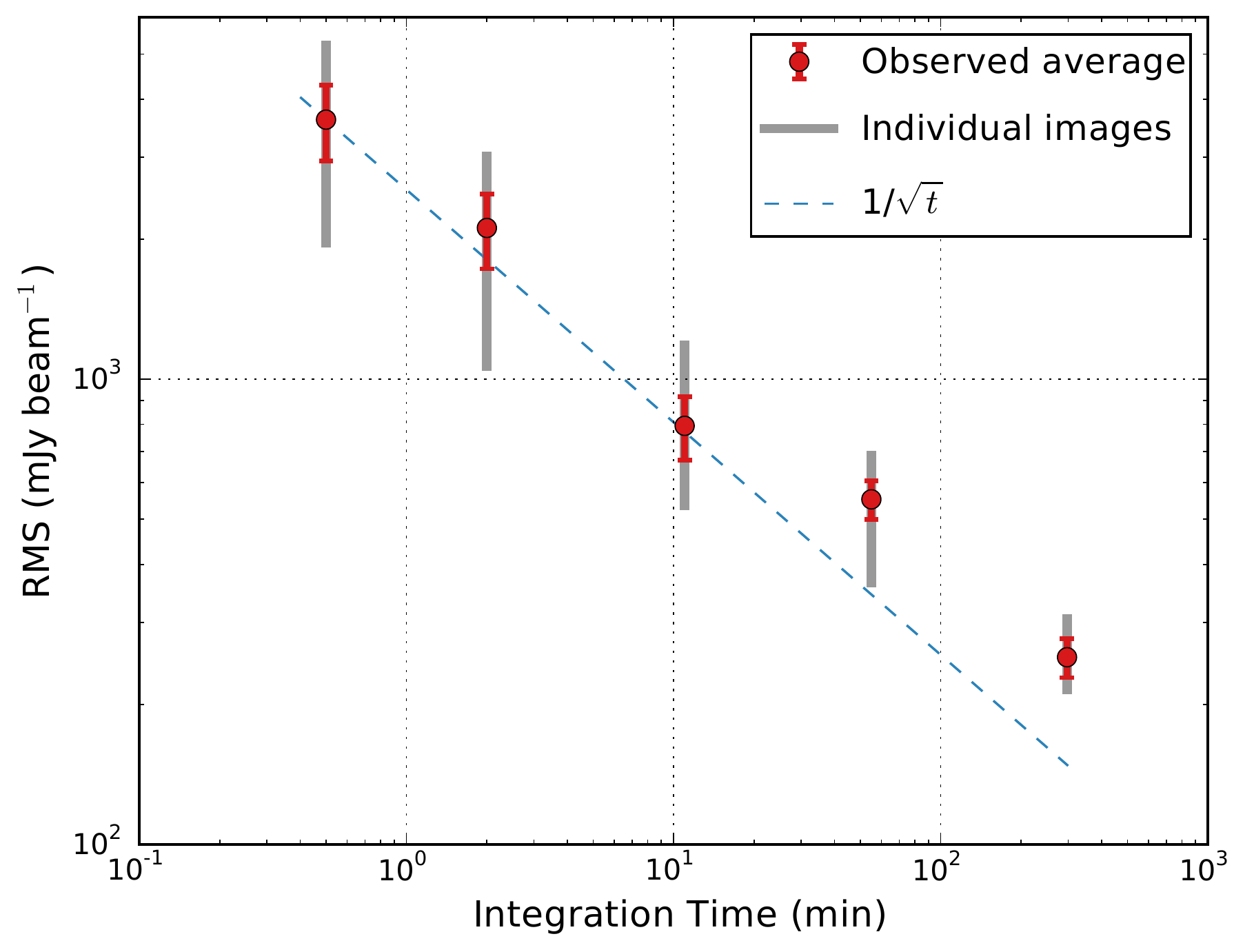}
\caption{The average rms obtained from the images produced by combining and splitting the dataset. Also plotted in light grey are the range of noise values for the individual images at their respective time-scales, in addition to the $1/\sqrt{t}$ relation where $t$ is the integration time of the observation. It can be seen that the average rms values approximately follow this relation; the longer time-scale values are likely to be slightly overestimated due to the methods used to estimate the rms. The errors shown on the average points are one standard deviation of the rms measurements from the respective time-scale.}\label{fig:rootn}
\end{figure}

It is important to note that, as a consequence of the primary beam correction, search areas centred on the NCP do not have a uniform noise level. Larger search areas include noisier regions further from the phase centre, and hence the flux density threshold at which we could detect a transient across the full search area will be higher. Figure~\ref{fig:ncpPB} shows an example of a primary beam map from one of the NCP observations. In order to obtain a noise estimate accounting for the variation caused by the beam, for each image at each time-scale we split the area into four annuli, equally spaced in radius. These four regions are also marked on Figure~\ref{fig:ncpPB}. The rms for each annulus was then measured, using a clipping technique, with the area-weighted average of these four values providing the single value rms estimate for the individual image. We then took the average of each time-scale, which are used as our sensitivity levels in Table~\ref{table:times}. Figure~\ref{fig:rootn} shows that these measured rms values of the different time-scales approximately follow a $1/\sqrt{t}$ relation, where $t$ represents the integration time of the observation. We note that the longer time-scale rms values appear to lie above the $1/\sqrt{t}$ relation. We believe this is caused by the clipping technique being less accurate at measuring the rms of the longer time-scale images annuli. This in itself due to the presence of many more sources compared to the relatively source free short time-scale images. In addition to this, it is possible the CLEAN algorithm was not applied to a deep enough level in some cases. Hence, the combination of these two methods means that the longer time-scale rms values are likely to be slightly overestimated, but not at a concerning level in the context of this investigation.

We could have limited the transient search to a smaller region with the deepest sensitivity; however, when calculating the figure of merit (FoM, $\propto \Omega s^{-\frac{3}{2}}$ where $\Omega$ is the FoV and $s$ is the sensitivity) it can be shown that it is more beneficial to extend the area of the search, despite the increase in average rms. This can easily be demonstrated as the full area is 16 times larger but the weighted sensitivity only drops by a factor of about two; hence the FoM is around five times better, illustrating the motivation for searching wide area. We refer the reader to \citet{FoM} for an in-depth discussion of the FoM in the context of transient surveys.

\begin{figure}
\centering
\includegraphics[scale=0.48]{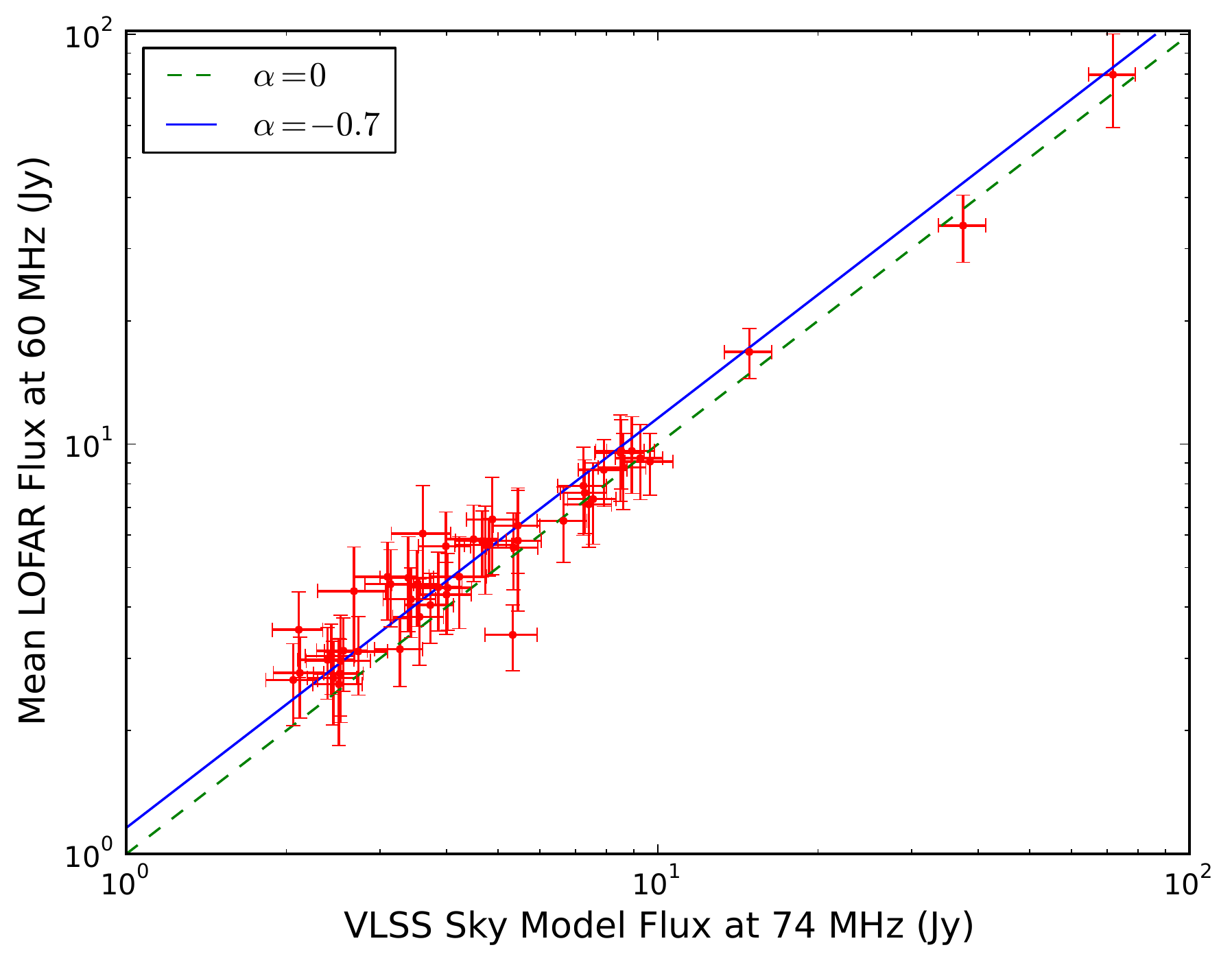}
\caption{Plot of the mean extracted flux of sources from the 297 minutes NCP survey at 60 MHz against the cross-matched VLSS survey at 74 MHz. The solid line represents the expected LOFAR flux density assuming a spectral index of $\alpha=-0.7$. For illustrative purposes a dashed-line representing $\alpha=0$ (a 1:1 ratio) is also shown.}\label{fig:vlssflux}
\end{figure}

The 55 and 297 min time-scale images offered the best flux calibration stability from image to image due to the better \textit{uv} coverage achieved on these time-scales. An example of the general flux calibration quality can be seen in Figure~\ref{fig:vlssflux}, which shows the averaged measured flux across all the 297 minute snapshots of sources detected at 60 MHz, cross-matched with the VLSS catalogue at 74 MHz. It shows a general agreement with the fluxes that would be expected assuming an average spectral index of $\alpha=-0.7$. If we assume that all sources have this spectral index and calculate the expected VLSS 60 MHz flux for each source, we find that the average ratio of this expected VLSS flux against the measured LOFAR flux is $1.00 \pm 0.17$.

\begin{table}
\centering
\caption{The average image sensitivity and number of epochs for each time-scale at which a transient search was performed. The accepted epochs column defines how many of the total number of images passed the {\sc trap} image quality control.}
\label{table:times}
\begin{tabular}{|c|c|c|c|c|}
\hline
Time & Average rms&Typical Resolution&Total \# & Accepted \# \\
(min)&(mJy beam$^{-1}$)&(arcmin) &Epochs&Epochs\\
\hline
\hline
0.5 & 3610 &4.8 $\times$ 2.2&47\,970&41\,340\\
2 & 2110 &4.7 $\times$ 2.1&10\,739& 9\,262\\
11 & 790 &5.4 $\times$ 2.3& 2\,149&1\,897\\
55 & 550 &4.9 $\times$ 2.1&371&328\\
297 & 250 &3.1 $\times$ 1.4&34& 32\\
\hline
\end{tabular}
\end{table}

Overall there was a typical scatter of 10 per cent in each light-curve of sources detected, which was measured by the {\sc trap}. It was common that fainter sources ($<10\sigma$) would appear to `blink' in and out of images; this was especially apparent in the 11 minute snapshots. This was likely due to a mixture of varying rms levels and the ionosphere causing phase calibration issues. Such behaviour was a further reason why a 10$\sigma$ source detection limit was used in the transient search. The sensitivities of the shortest time-scale maps, 30 s and 2 min, were such that only the brightest source in the field, 3C\,61.1, was detectable. The LOFAR and VLSS source positions were also consistent within 5.1 arcsec on average; the typical resolution in the LOFAR band is 3.1 $\times$ 1.4 arcmin for the 297 min time-scale.\\

It was also important to determine whether the images produced for the transient search are confusion limited. In order to calculate an estimate of the confusion noise for the average resolutions presented in Table~\ref{table:times}, we followed the same approach as \citet{MSSS}, using VLSS C-configuration estimates \citep[see][]{confusion} which we extrapolate to 60\,MHz using a typical spectral index of $-0.7$. We also alter the formula to account for the non-circular beams:
\begin{equation}	
\label{eq:confusion}
\sigma_{\textrm{conf,VLSS}} = 29\left(\frac{\theta_{\textrm{1}}\times\theta_{\textrm{2}}}{1^{\prime\prime}}\right)^{0.77}\left(\frac{60\,\textrm{MHz}}{74\,\textrm{MHz}}\right)^{-0.7} \ \umu \textrm{Jy beam}^{-1} \textrm{ ,}
\end{equation}
where $\theta_{1}$ is the synthesized beam size major axis and $\theta_{2}$ is the minor axis. For the five time-scales used in the transient search shown in Table~\ref{table:times}, beginning with 30 sec, we calculate the confusion noise estimates to be 113, 107, 128, 111 and 57 mJy beam$^{-1}$ respectively. Thus, due to our simple reduction strategy, our images, at best, are approximately $4\times$ the confusion noise level and hence would not affect our transient search.\\

\begin{figure}
\centering
\includegraphics[scale=0.31]{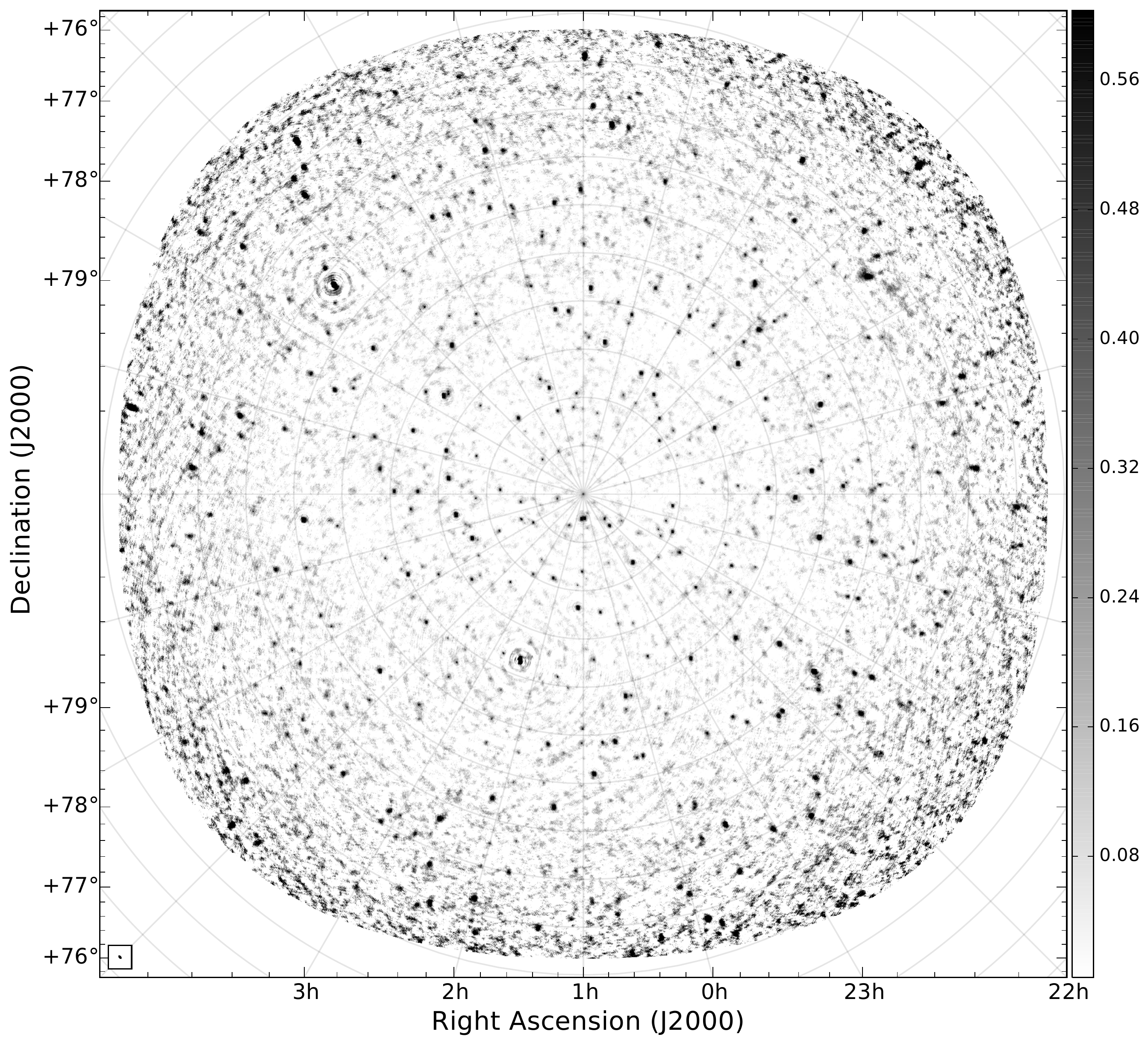}
\caption{The deepest map produced of the NCP field from the survey. It was constructed by averaging all 31 of the 297-minute-duration images together in the image plane, using inverse-variance weighting. It has a noise level of 71 mJy beam$^{-1}$ and a resolution of $3.1 \times 1.4 \textrm{ arcmin}$ (BPA $42^\circ$). A total of 150 sources are detected at a 10$\sigma$ level within a radius of 7.5 deg from the centre of the map. While none of these sources are previously undetected, it provided a detailed reference map to check any transient candidates. The colour bar units are Jy beam$^{-1}$.}\label{fig:deep}
\end{figure}

Along with these cadences, a deep map was constructed by using all the available 297 min images, reaching a sensitivity of 71 mJy beam$^{-1}$ (this value was measured using the weighted average method discussed above in this section). This map can be seen in Figure~\ref{fig:deep}. This, however, had to be produced by means of image stacking as opposed to direct imaging due to the amount of data involved. A total of 150 sources were detected at a $10\sigma$ level within the same 7.5 deg radius circle used for the transient search, with the map primarily being used as a deep reference image for the field. We can, however, use this deep map to verify our calibration and imaging procedures by comparing our detected source counts to the VLSS. Firstly, using a spectral index of $-0.7$, $S_{60}=710$ mJy corresponds to a flux density at 74 MHz of $S_{74}=613$ mJy. Using this flux density limit, there are 263 catalogued VLSS sources within 7.5 deg of the phase centre. Cross-correlating the VLSS with our LOFAR 60 MHz detections, we find that 41 per cent of the VLSS sources have a LOFAR match. The factor of $\sim2$ discrepancy can be shown to be simply due to the primary beam attenuation in our deep map. Hence, we were satisfied that the calibration and imaging results were valid and consistent with previous studies, and therefore would not negatively impact any transient searches.

This map was also further analysed for any previously uncatalogued radio sources, but none were found. However, the direct comparison to VLSS revealed that one source, located at $02^{\mathrm{h}}13^{\mathrm{m}}28^{\mathrm{s}}$ +$84^{\circ}04^{\prime} 18^{\prime\prime}$, has apparently significantly different 60 and 74 MHz flux densities: the VLSS integrated flux density is 1.49 Jy (possibly put in the error), whereas in the LOFAR band it is detected at the 8$\sigma$ level with a integrated flux density of 236 mJy. There are no detections of the source in WENSS or NVSS. However, this source is located within a stripe feature in the VLSS image, and the source is not present in the VLSS Redux catalogue (VLSSr; \citealt{vlssr}); hence we do not pursue this source further. The full MSSS survey will offer further insight into this potential source, confirming its flux density and spectral index, if it is real.

\begin{figure}
\centering
\includegraphics[scale=0.27]{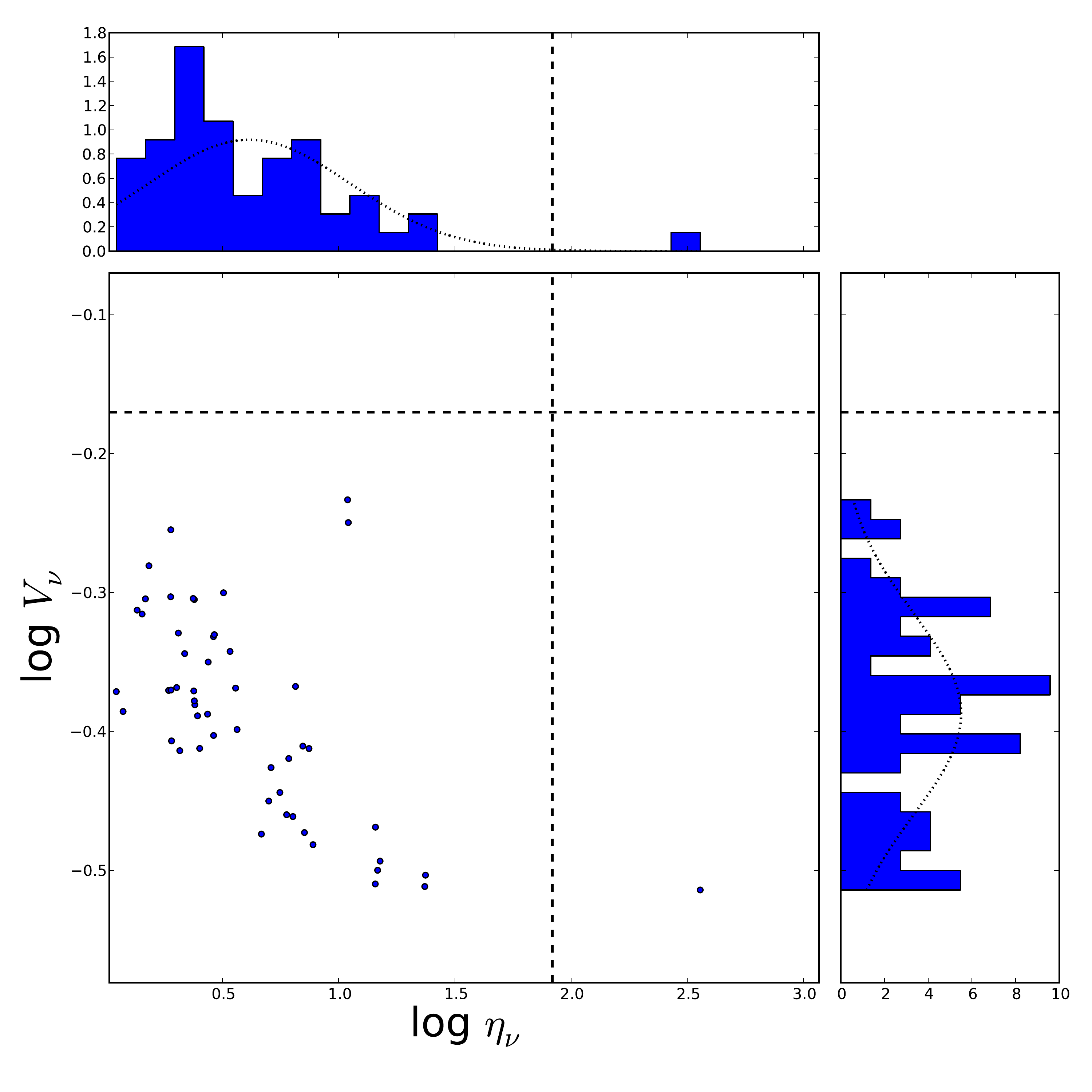}
\includegraphics[scale=0.27]{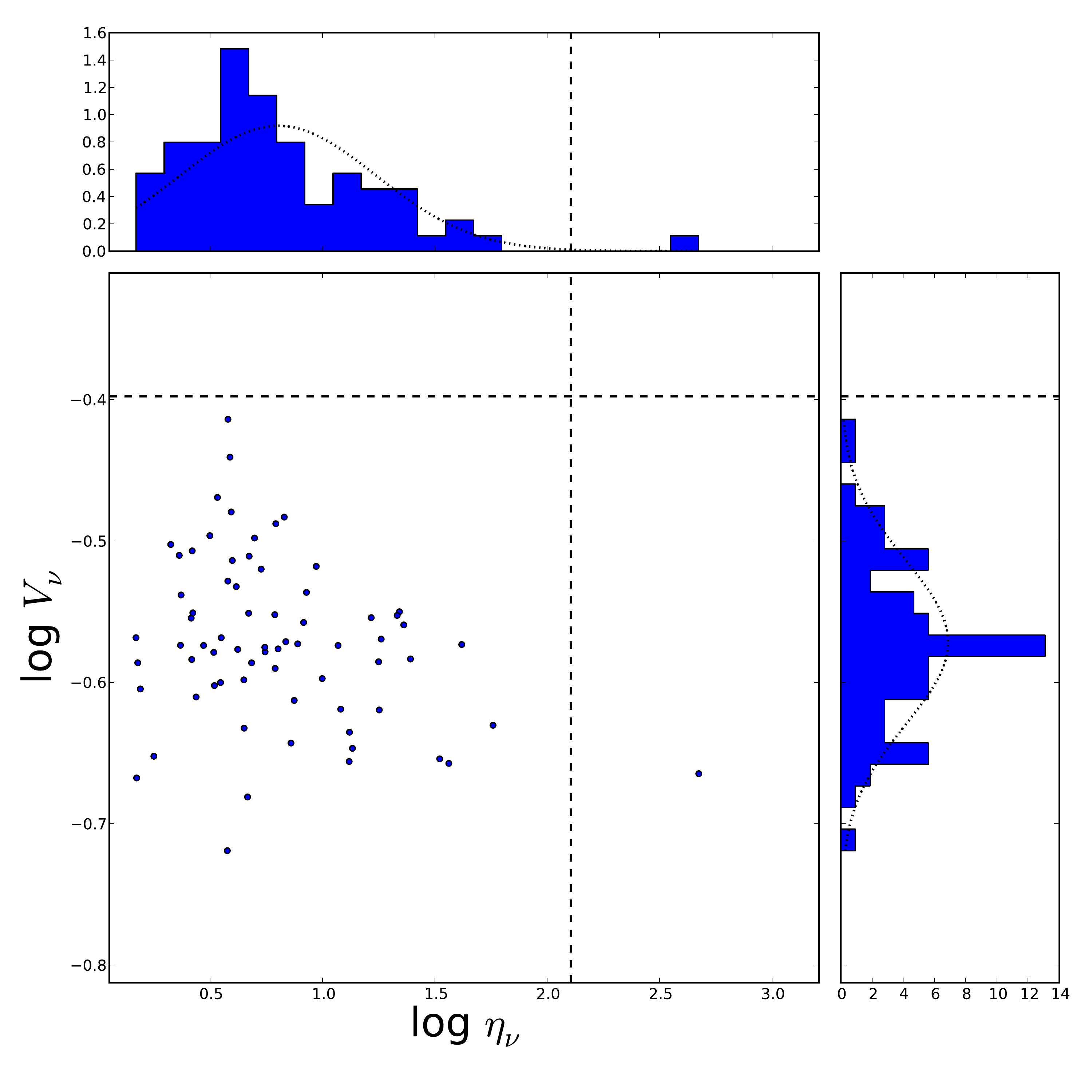}
\caption{This figure shows the distribution of values obtained for the variability parameters $V_\nu$, a coefficient of variation, and $\eta_\nu$, the significance of the variability (see text for full definitions) for each light-curve detected. The \textit{upper panel} shows the 55-minute image results and the \textit{lower panel} shows the 297-minute time-scale results. In each case, the central panel plots the two values against each other for each source, with the top panel and right side panel displaying the histogram showing the distribution of the $\eta_\nu$ and $V_\nu$ values respectively for all sources. The dotted lines represent a $3\sigma$ threshold for each parameter. A very-likely variable or transient source would appear in the top-right of the plot, exceeding a $3\sigma$ level in each parameter. At both time-scales, one source (3C\,61.1) is found to have a significant value in $\eta_\nu$. However this is likely to arise from fluctuations caused by calibration issues.}\label{fig:ncpvar}
\end{figure}

\subsection{Variability search results}
\label{sec:varresult}
The four month dataset provides an opportunity to search for variable sources as well as transient sources. We define variables as sources which are present throughout the entire dataset, taking into consideration varying sensitivity, whose light curve displays significant variability over the period. This is opposed to transient sources, which we define as sources that appear or disappear during the time spanned by the dataset, again taking into account the varying sensitivity. Consulting historical catalogues also helps with the distinction between variables and transients. Due to the higher level of image quality, the variability search was limited to the two longest time-scales of 55 and 297 min. For each detected source in these two sets of images, variability parameters ($V_\nu$ and $\eta_\nu$) were calculated by the {\sc trap}. Figure~\ref{fig:ncpvar} shows the respective distributions of the variability parameters for each time-scale plotted in logarithmic space. In each case, the central panel shows $\eta_\nu$ plotted against $V_\nu$ for each detected source. The top panel displays a histogram representing the distribution of $\eta_\nu$ of all the sources along with a fitted Gaussian curve. The right panel contains the distribution and fitted Gaussian curve for $V_\nu$. The dashed lines represent a 3$\sigma$ threshold for each value; any sources with variability parameters exceeding one or both of these values are considered as potentially variable. Candidates also had to show a variability of significantly more than ten per cent, which was the calibrator error of the measurements. This was set at a level of $2\sigma$ from this value. An ideal transient would appear in the top-right-hand corner of the central panel scatter plot, exceeding the threshold in each parameter.\\

It can be seen that at both time-scales, no sources exhibit variable behaviour in $V_\nu$ above a 3$\sigma$ level, but one source has a significant $\eta_\nu$ value. This source is 3C\,61.1, which dominates the field. While the result points towards low-level variability of 3C\,61.1, the source is a well resolved radio galaxy \citep{3C61} whose flux is dominated by 100-kpc-scale lobes, making it very unlikely that we would detect any intrinsic variability. It is more likely that this is the result of calibration errors and the source extraction and subsequent calculation of $\eta_\nu$ itself. The model for 3C\,61.1 used during this investigation is quite basic for such a complex source. This, along with ionospheric effects and the general calibration accuracy of the instrument at the time, can have quite a substantial effect on such a bright source, with such calibration errors not included in this analysis. The source is also spatially extended, but the extraction treats it as a point source (as mentioned in Section~\ref{sec:trap}), and this will therefore also have a significant impact on the recorded flux. Removing the point source fitting constraint does indeed move the data point closer back towards the 3$\sigma$ threshold, but only marginally by $0.1$ dex in $\eta_\nu$. As for the $\eta_\nu$ value, this parameter is weighted by the flux errors of the source extraction. Bright sources, such as 3C 61.1, are well fitted when they are extracted, which means they have small associated statistical flux errors. This in turn then causes $\eta_\nu$ to rise. If we discount 3C\,61.1, no sources displayed any significant variability at the 55 and 297 minute time-scales.

\subsection{Transient search results}
\label{sec:transresult}
Using the {\sc trap} and a manual analysis of its results, searches performed on the time-scales of 0.5, 2, 55 and 297 minutes found no transient candidates. However, nine transient candidates emerged from the analysis of the 11-minute time-scale. At first it appeared strange to achieve nine candidates at one time-scale but none at any other. However, the sensitivity of the shorter time-scales was such that only bright transients ($> 25$ Jy) would have been confidently detected, and as previously stated no other source, or even artefact, was detected at these flux levels other than 3C 61.1. At the longer time-scales, the improved \textit{uv} coverage meant that the images improved substantially in quality. This reduced the number of imaging artefacts that could spawn false detections and sources were consistently detected throughout the epochs (as opposed to many sources blinking in and out as discussed in Section~\ref{sec:imagequality}). Any sources that were defined as `new' by the {\sc trap} (sources which appeared that were not detected in the first image) were in fact association errors and not transient sources.\\

While the nine candidates could point towards the 11 minute images meeting the required sensitivity and time-scale of a transient population, these images are also the most likely to exhibit misleading artefacts due to the limited \textit{uv} coverage. Hence, the nine reported candidates were subjected to a series of tests to determine whether they were spurious sources. The following tests were performed:
\begin{enumerate}[leftmargin=*]
\item Subtraction of 3C\,61.1 from the visibilities using the clean component model from the deconvolution process. The visibilities were then re-imaged.
\item Applying an extra round of RFI removal using {\sc aoflagger}.
\item Re-running the automated tool to remove perceived bad LOFAR stations from the observations, followed by a manual check.
\item Imaging the data using different weighting schemes and baseline cutoffs.
\end{enumerate}
The tests were applied in the above order, meaning that if one method definitely succeeded in removing the candidate the latter tests were not performed. Only one of the nine candidates completely survived all the tests; three were inconclusive but quite doubtful, whereas four were definite artefacts. One other source was very marginal in passing all the tests; hence this event is not presented in this paper, but will be discussed in a future publication. The surviving candidate was thus a potential real astrophysical event and is the subject of the following Section~\ref{sec:transient1}.

\section{Transient Candidate ILT J225347+862146}
\label{sec:transient1}
The only candidate to have passed all the validity checks, was found in a single 11 min snapshot taken on 2011 December 24 at 04:33 \textsc{utc}. The source was extracted by the {\sc trap} with a flux of 7.5 Jy ($14\sigma$ detection in individual image), at coordinates $22^{\mathrm{h}}53^{\mathrm{m}}47.1^{\mathrm{s}}$ +$86^{\circ}21^{\prime} 46.4^{\prime\prime}$, with a positional error of $11^{\prime\prime}$. It was only seen in this one snapshot with no detection of the source in the preceding or subsequent snapshots. The observation can be seen in Figure~\ref{fig:ncp_ghosts}. Nothing was present at the candidate position in either the relatively deep image constructed from the longer time-scale images (see Section~\ref{sec:imagequality}) or the very deep image of the field from the LOFAR Epoch of Reionisation (EoR) group \citep{sarod}. Note that the EoR project uses the LOFAR high-band antennas, and hence it is at a higher frequency range of 115--163 MHz.

\subsection{A mirrored ghost source}
\label{sec:ghost}
\begin{figure}
\centering
\includegraphics[scale=0.55]{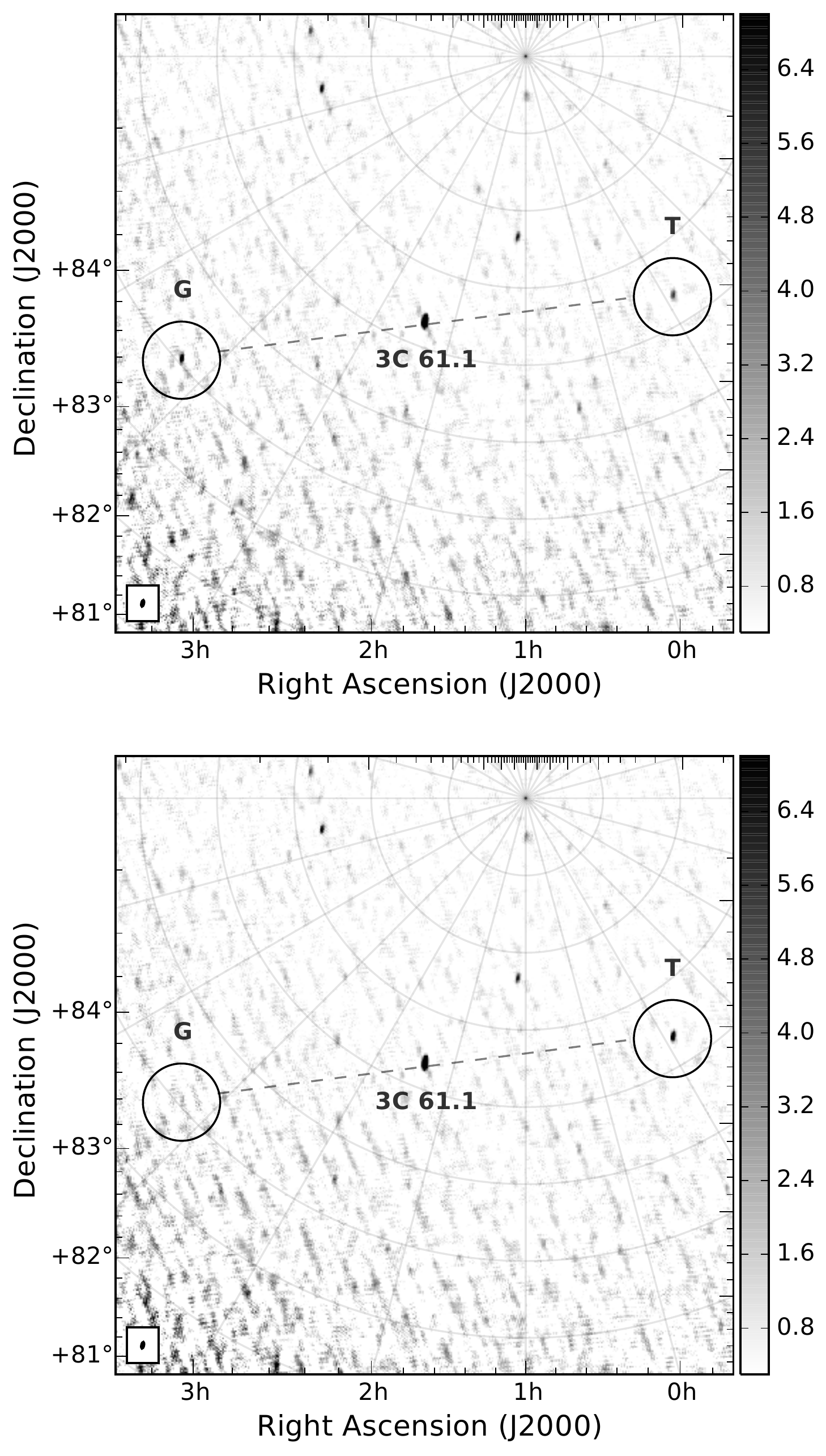}
\caption{\textit{Upper Panel}: Illustrates how the transient source (labelled `T'), ILT J225347+862146, was originally detected in the image, along with the associated ghost source (labelled `G') across from 3C\,61.1. \textit{Lower Panel}: Now the measurement set as been re-calibrated with the transient included in the sky model; the ghost source has vanished. Upon closer inspection, other faint, source-like features also disappear from the re-calibrated image. These are most likely fainter ghost features which are reduced when the data were calibrated with a more complete sky model. The colour bar units are Jy beam$^{-1}$.}\label{fig:ncp_ghosts}
\end{figure}

On closer inspection, the transient candidate appeared to have a secondary associated positive `ghost' source mirrored across the brightest source in the field, 3C\,61.1 (the transient lies at an angular distance of 3.2$^{\circ}$ from 3C\,61.1), which can also be seen in Figure~\ref{fig:ncp_ghosts}. This ghost was not detected by {\sc trap} due to the higher rms value in that region, and like the transient candidate it was a `new' source with no previous or subsequent detections. In fact the ghost source was actually nominally brighter than the transient source with a flux density of 13 Jy. However, in the non-primary-beam-corrected map the candidate has a higher peak flux density (9 Jy) than the ghost (6 Jy). This was not the first time we had witnessed this type of effect in LOFAR observations, with previous commissioning data we had obtained in 2010 showing a similar situation. Currently, the exact explanation of why ghosts of this nature, including specifically the ghost presented in this work, are generated in LOFAR data is unknown. It should be noted that none of the other eight transient candidates detailed previously had an associated ghost source. In the following discussions we refer to the original detected transient source ILT J225347+862146, to the west of 3C\,61.1, as the `transient candidate' and the source to the east of 3C\,61.1 as the `ghost' source (refer to Figure~\ref{fig:ncp_ghosts}).

\subsubsection{Ghost artefacts in radio interferometry}
\label{sec:ghostsinradio}
Calibration artefacts presenting themselves as spurious `ghost' sources is not an entirely new topic to radio interferometry. The topic of `spurious symmetrisation' is discussed in \citet{cornwell}; in brief, if a point source model is used for a slightly resolved source, a single iteration of self-calibration can result in features of the image being reflected relative to the point-like object. However, this can be corrected with further iterations of self-calibration which would cause the spurious features to disappear. As will be discussed in Section~\ref{sec:ncpghost}, the ghost presented in this work can be seen before initiating any kind of self-calibration of the target field, i.e. any calibration using a target field sky model. Therefore, it is highly unlikely that the spurious symmetrisation previously described is the sole cause of the ghost. However, this is not to say that the effect plays no role in its creation.\\

More recently, \citet{ghost} (hereafter `G14') began a series of investigations dedicated to ghost phenomena. This first study concentrated on ghosts seen in data from the Westerbork Synthesis Radio Telescope (WSRT). In these data, ghost sources appeared as strings of (usually) negative point sources passing through the dominant source(s) in the field. The arrangement of these negative point sources appeared quite regular, along with the fact that the positions were not affected by frequency. In their investigation, G14 were successful in deriving a theoretical framework to predict the appearance of ghosts in WSRT data for a two-source scenario, and were able to confirm what previous work had suggested concerning these ghost sources (see text in G14).

In brief, the main features about ghosts to note are as follows:
(i) they are associated with incomplete sky models, for example missing or incorrect flux;
(ii) in the WSRT case, the ghosts always formed in a line passing through the poorly modelled or unmodelled source(s) and the dominant source(s) in the field;
(iii) the ghosts are mostly negative in flux, while positive ghosts are rare and weaker; and
(iv) the general ghost mechanism can also explain the observed flux suppression of unmodelled sources.

G14 also concluded that the simple East-West geometry of the WSRT array is the reason for ghosts appearing in a regular, straight line, pattern. This becomes more complex when a fully 2D/3D array is considered such as LOFAR, where the ghost pattern is expected to become a lot more scattered and noise-like. This subject will be the focus of Paper II (Wijnholds et al. in prep.) in the series on ghost sources. However, G14 did note that regardless of the array geometry, ghosts are expected to occur at the $n\phi_0$ positions, where $\phi_0$ represents the angular separation between the respective bright source and unmodelled source, and $n$ is an integer number. Usually the strongest ghost responses are the $n=0$ and $n=1$ positions, i.e. the suppression ghosts that sit on top of the sources in question. However the case discovered in this work, and also two independent cases (de Bruyn, priv. comm., Clarke, priv. comm.) in LOFAR data suggest that the $n=-1$ position could also generate a strong response. What is significant about the transient presented in this work, however, is that the ghost appears as a positive source.

\subsubsection{Investigating the NCP ghost}
\label{sec:ncpghost}
\begin{figure}
\centering
\includegraphics[scale=0.37]{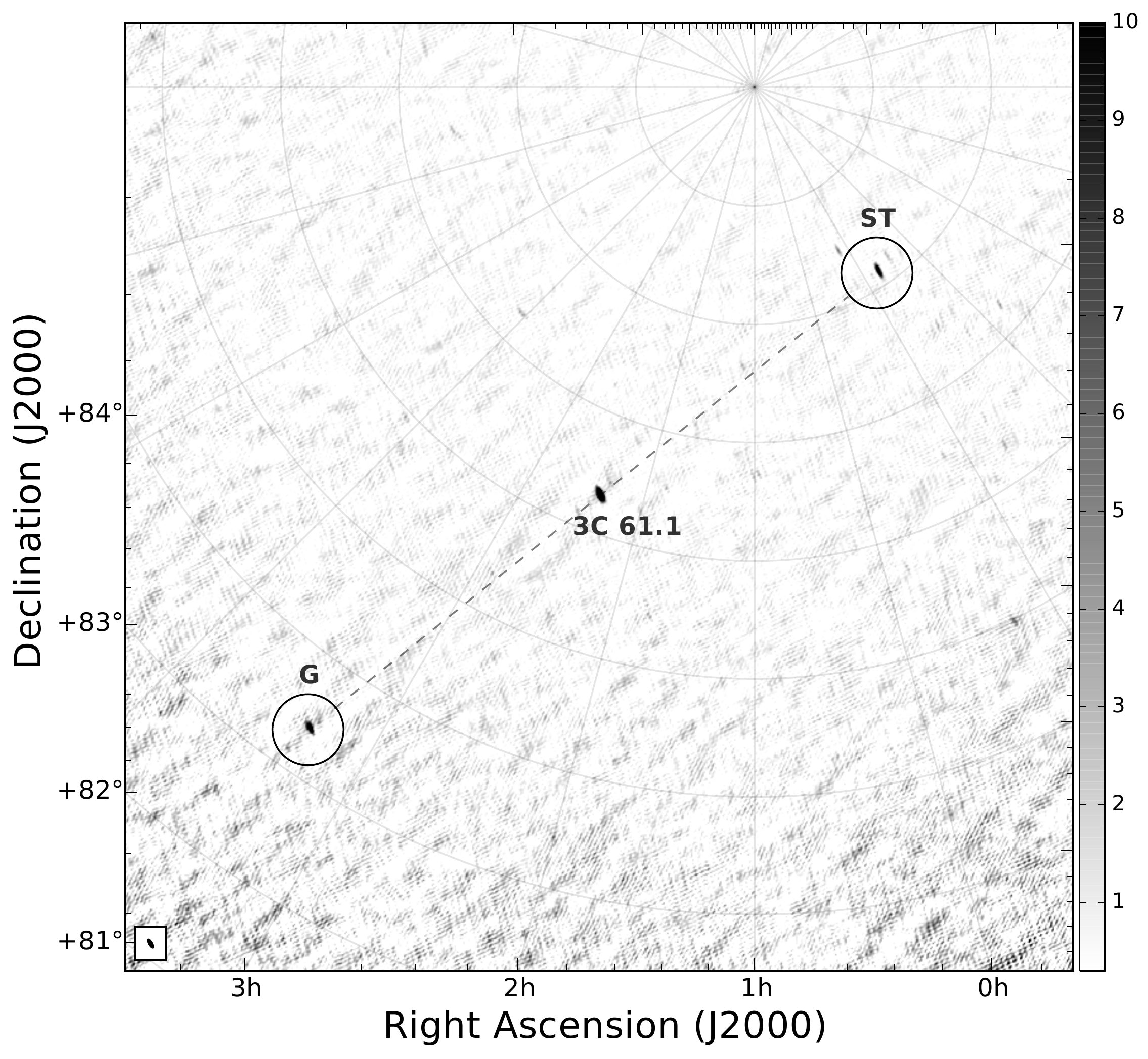}
\caption{The resultant image after a simulated transient source (labelled `ST') was inserted into the visibilities of a NCP observation and processed without the simulated source in the sky model. A ghost source (labelled `G') appears mirrored across 3C 61.1. The effect is not limited to one specific insertion point of the simulated transient and is more pronounced the brighter the simulated transient. In this example, a source of brightness 80 Jy was inserted, which produces a very significant ghost source. The simulated transient and ghost source each had a measured flux density of $\sim$25 Jy, with the remaining $\sim$30 Jy being absorbed by 3C\,61.1. This transfer of flux was common when the simulated transient was brighter than 3C\,61.1 ($\sim$80 Jy). When lower, the flux is shared equally between the simulated and ghost sources, with minimal flux transferred to 3C\,61.1. The colour bar units are Jy beam$^{-1}$.}\label{fig:simghost}
\end{figure}
Returning to the situation detailed in this paper, we were presented with two sources for which either could be the real (transient) source or the ghost. We attempted to simulate the situation within real data, in order to investigate how the different stages of calibration would react to a bright transient, and if we could also generate a positive ghost source. This was done by taking a different NCP observation and inserting a simulated transient source into the visibilities (the transient was set to be `on' for the entire 11 mins) before any calibration had taken place. The snapshot was then calibrated as normal, but importantly the inserted source was \emph{not} included in the NCP sky model used for the phase-only calibration step (refer to Section~\ref{sec:imaging}). This test was repeated using various different sky positions and flux densities for the inserted source. We found that we could produce a significant positive ghost source only if the flux of the simulated transient was relatively bright, $\sim$40 Jy. An example can be seen in Figure~\ref{fig:simghost}. We observed that it was common for the total flux to be shared approximately equally between the simulated source and its associated ghost. However, not every position on the sky at which the transient was inserted produced a ghost source, a feature that we cannot currently explain. Yet, when a transient was inserted at the position of ILT J225347+862146, this did produce a ghost source. We were then able to test what happened when the simulated source was included in the sky model. We observed that when the simulated source was accounted for perfectly in the sky model, the ghost source disappeared. If the sky model component was instead inserted at the location of the ghost source, while the ghost appeared brighter, the simulated transient never fully disappeared.\\

In light of the results from the simulations, we performed the same sky model test with the transient candidate and ghost in order to determine which source was the `real' source. Recalling that the total flux of the transient candidate and ghost was $\sim$7 Jy + $\sim$13 Jy $\approx$ 20 Jy, we began by inserting a 20 Jy point source into the NCP sky model at the position of the transient candidate and re-calibrated the dataset. We found that in this case the flux of the ghost was significantly reduced, by $\sim$70 per cent, and the candidate brightened by $\sim$100 per cent. Alternatively, if the model component was entered at the ghost location, the candidate source and ghost respective fluxes were only $\sim$10 per cent different from their initial fluxes on discovery, i.e. when they were not in the sky model at all. In fact increasing the sky model component to 25 Jy and placing it back at the position of the transient candidate reduced the ghost such that it was no longer distinguishable from the noise, as seen in the bottom panel of Figure~\ref{fig:ncp_ghosts}. Hence, the `real' source was determined to be at the position {\sc trap} had originally reported, $22^{\mathrm{h}}53^{\mathrm{m}}47.1^{\mathrm{s}}$ +$86^{\circ}21^{\prime} 46.4^{\prime\prime}$, to the west of 3C\,61.1.\\

The above tests have concentrated on the target NCP field sky model, but we also have the sky model which was used to calibrate the calibrator observation. For this observation, the calibrator source was 3C\,295. Considering that ghosts occur because of sky model errors, one could envision a scenario in which the error being transferred from the calibrator to the target field results in the ghost pattern observed. As mentioned in Section~\ref{sec:imaging}, the calibrator sky models only contain the calibrator source itself and not any surrounding field sources. While this generally allows the derivation of sufficiently accurate gain solutions, the missing flux could be attributed to a ghost pattern, which is then transferred to the target field (see also \citealt{3C196Calib} for a similar discussion regarding the 3C\,196 field).

To investigate this, two tests were performed. Firstly, the phase-only calibration step was ignored and we imaged the dataset using the amplitude and phase gain solutions directly from the calibrator. In this case, both the transient and the ghost were present, with no major changes from before (a result which makes `spurious symmetrisation', previously discussed in Section~\ref{sec:ghostsinradio}, unlikely to be the sole cause of the ghost). Secondly, the calibrator observation was not used at all and instead the data were calibrated in both amplitude and phase using the constructed NCP target sky model (described in Section~\ref{sec:imaging}) which importantly did \emph{not} contain the transient source. For this test, we increased the solution interval to one minute (originally 10 s) to gain more signal-to-noise for the calculations. We also had to perform post-processing clipping to the visibilities to eliminate bad amplitude spikes in the calibrated visibilities. In the full 11 min image, while the rms rose to $\sim$800 mJy beam$^{-1}$, a source was detected within one arcmin (the resolution of the image was $5.6$ $\times \ 2.4$ arcmin) of the reported transient candidate position with a flux density of 13 Jy. The ghost source was not detected to a $5\sigma$ limit of 10 Jy at its expected location, nor was it visible when the map was manually inspected. However, due to the increase of the rms in this case, we cannot state with complete confidence that the ghost source is not present at all. Nonetheless, observing the transient source without placing it in the sky model provided additional evidence that we had identified the correct source.\\

The above result tentatively points to the calibrator having an important role in the ghost creation. However, understanding the exact ghost mechanism is a complex task in the LOFAR case, and each stage of the calibration must be taken into careful consideration. For example, G14 has exclusively investigated situations where full amplitude and phase calibration is used, so the effects of a phase-only calibration is generally unknown at this stage. At the time of writing, we cannot explain how the ghost is generated; a detailed investigation is under way (Grobler et al. in prep.) to resolve the matter.

\subsection{Transient flux density}
\label{sec:flux}
\subsubsection{Obtaining the correct flux}
\label{sec:correctflux}
\begin{figure}
\centering
\includegraphics[scale=0.23]{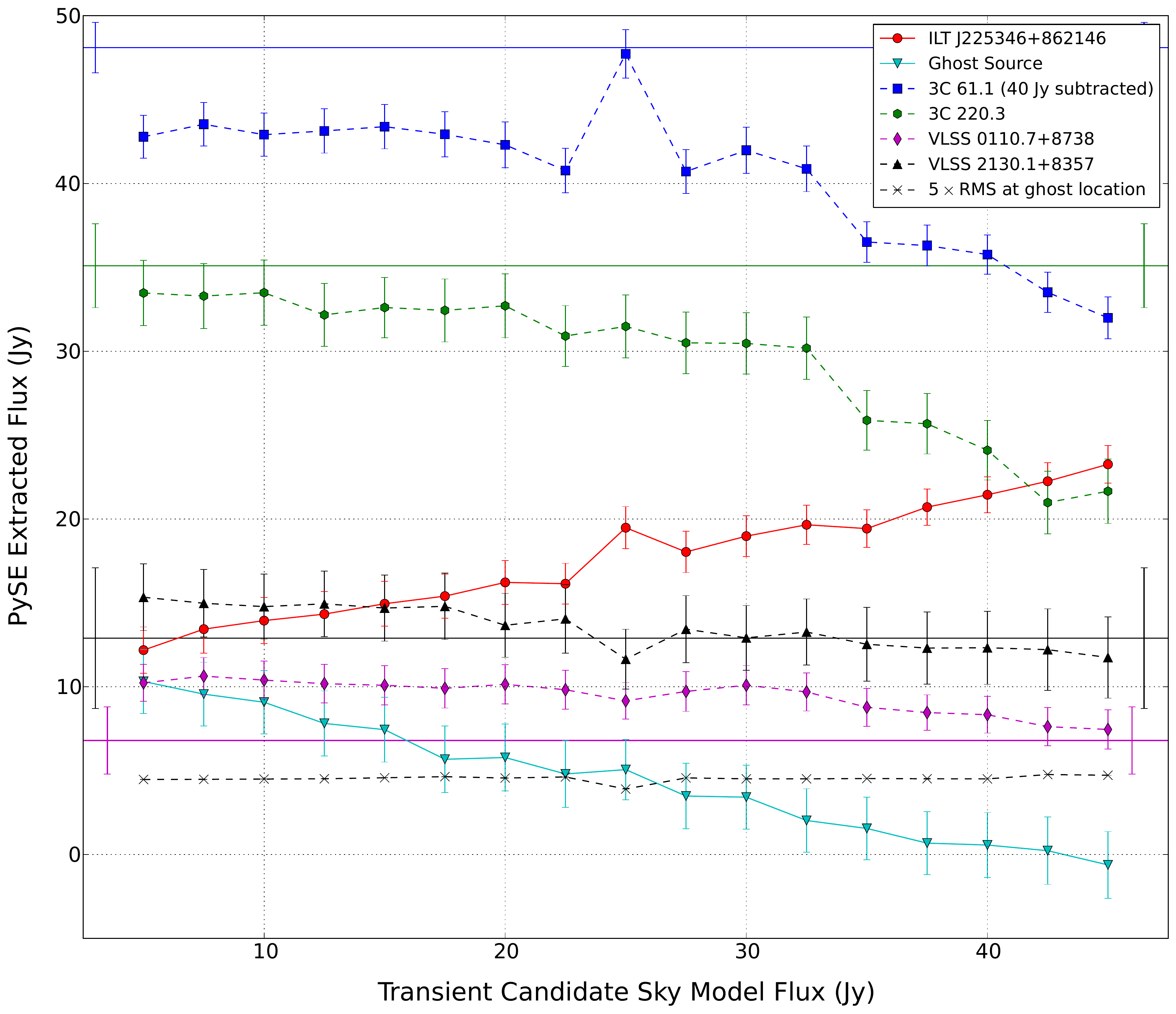}
\caption{The extracted flux of the transient candidate using the PySE source extractor against the manually defined flux entered into the sky model at the transient position when processing. Also shown is a measurement of the ghost source flux obtained by a forced fit at the ghost position. The input flux was defined in steps of 2.5 Jy, from 7.5 Jy to 45 Jy. The plot also shows the extracted fluxes of four other sources in the field in order to monitor any effects to other sources, along with the solid lines which show the average flux of these field sources from the four surrounding snapshots. The error on these averages is shown by the error bar at the beginning and end of the line. Above an input flux value of 20 Jy (extracted transient flux value of 17 Jy) it becomes apparent that the other sources are beginning to be affected. They drop sharply beyond an entered flux of 30 Jy by which point the ghost source is no longer statistically significant. Note that 3C 61.1 has been scaled by subtracting 40 Jy from its flux measurements.}\label{fig:fluxes}
\end{figure}

The correct flux density of the transient proved difficult to ascertain. We attempted to obtain an estimate by entering flux values of the transient source manually into the calibration sky model, over a range of 7.5--45 Jy, in steps of 2.5 Jy, and proceeded to recalibrate the visibilities (as previously described, this calibration step is phase only). We then observed the influence this had on the measured flux density of the transient itself, as well as the measured flux densities of the surrounding sources, including the ghost source. We remind the reader that the transient candidate, the source deemed `real', is to the west of 3C\,61.1 and the ghost is the source to the east of 3C\,61.1. The transient was always placed as a point source in the sky model. The results of this experiment can be seen in Figure~\ref{fig:fluxes}. We found that the ghost source became increasingly fainter as the transient flux was increased, right up until the transient was entered as 20 Jy and the ghost could no longer be distinguished from the background. The transient `light curve' itself follows the trend of the increasing sky model flux, but it also exhibits a sudden local maximum when the sky model entry level is changed from 22.5 to 25 Jy. In this instance the extracted flux rises from 16 Jy to 20 Jy. It then proceeds to fall back to an extracted flux level of 18 Jy and continues to rise as before.

As for the other nearby sources, while they are stable prior to the sky model transient component reaching 17.5 Jy, beyond this level they suffer a very noticeable decline that continues as the transient flux is increased. It is also apparent that the other sources in the field are affected by the before mentioned sudden local maximum of the transient light curve around a sky model flux of 25 Jy, with 3C\,61.1 also showing a significant flux increase ($\sim3\sigma$ to the scaled value). However, for VLSS 0110.7+8738 and VLSS 2130.1+8357, which are at a similar flux level to ILT J225347+862146, there is a hint of a decrease, although within the error bars of the flux measurements. In each case, once the sky model flux is increased to the next step, the measured fluxes return to their previous levels. When comparing the fluxes of the field sources with the corresponding averages from the four surrounding snapshots, we see that they mostly agree within all the error bars involved. The largest discrepancy comes from 3C 61.1, which appears $\sim10$ per cent dimmer in the transient snapshot, which is outside the errors of the average measurement. However, the sudden increase around 25 Jy causes 3C\,61.1 to match the surrounding average. This could be seen as a clue that this area represents the real flux of the transient; at this point, with 25 Jy in the sky model, the transient appears as 20 Jy in the image. Hence, with this information, we associate the true flux of the source with the point at which the ghost disappears and the other sources in the field are not heavily affected, which constrains our estimate of the flux density of ILT J225347+862146 to be in the range 15--25 Jy.

\subsubsection{Testing known sources}
The test detailed above was performed directly on the two field sources that were monitored during the investigation, VLSS 0110.7+8738 and VLSS 2130.1+8357, with 60-MHz flux densities of $\sim$9 and $\sim$15 Jy respectively. This also included removing the sources from the calibration sky model as well as changing the input flux. Each source was treated as a separate case meaning that both were never subtracted from the sky model, or edited, at the same time. As before, these tests were performed at the transient epoch, but also in the two neighbouring epochs to ensure that any effects were not just local to the transient-containing snapshot. Without the source in the model, the measured flux was reduced by $\sim$20 per cent, with the majority of the extra flux in the field being absorbed by 3C 61.1, which appeared slightly brighter. Once the source was reinserted into the sky model, even at a low flux, the source in question returned to the expected level. However, as the sky model input flux was increased, so did the extracted flux, which is consistent with how the transient acted previously. There was also no distinguishing feature that would enable a confident definition of these sources' `correct flux' without prior knowledge. Thus, it is not a surprise that the transient flux in Section~\ref{sec:correctflux} is hard to identify purely from the behaviour of the source itself during calibration when altering the sky model. Ideally self-calibration would be used, but at the time of processing self-calibration with LOFAR was still a relatively untested technique. 

\begin{figure}
\centering
\includegraphics[scale=0.47]{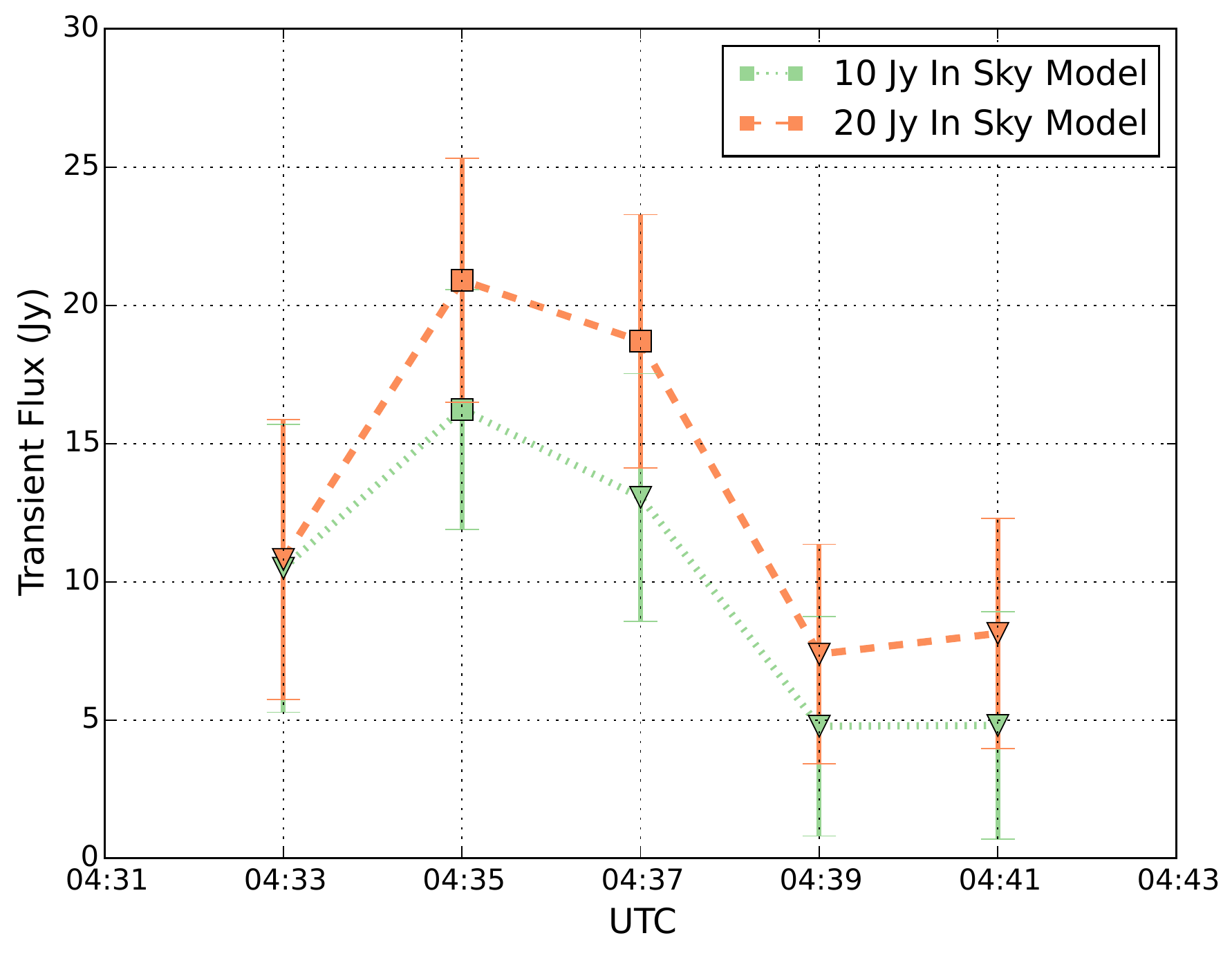}
\caption{The extracted flux of the transient candidate in 2-minute intervals, obtained using the {\sc trap}, for the cases where the transient is included in the sky model at 10 and 20 Jy. Data points represented by a square signify that the source was extracted from the image with a blind detection. In contrast, the triangles represent flux values obtained from a forced fit at the source position, where the source was no longer above the source extraction threshold (13 Jy, $5\sigma$). The two light curves follow the same trend, suggesting that the transient is brighter than a 3$\sigma$ limit of 7.5 Jy in the 4-minute period of 04:34 - 04:38. The 10-Jy input case, returning a measured flux of 16 Jy, also suggests that 16 Jy may be the correct flux during this time frame. These light curves were obtained by extending the phase-calibration time interval to one min. The date of the observation was 2011 December 24.}\label{fig:2minfluxes}
\end{figure}

\subsubsection{Splitting the dataset in time}
\label{sec:timesplit}
In the test detailed above, where the transient was inserted into the sky model with various different flux values, it was noticeable that the flux that was inserted was never the flux that was measured. If the transient was not `on' for the entire 11 min, this could perhaps explain why this was the case. With the transient included in the sky model at a flux of 20 Jy, the observation was firstly split in half and imaged; however, the flux was consistent within the $1\sigma$ error bars between each half. To probe deeper, we then referred to the 2-minute images produced as part of the transient search, which did not have the transient included in the sky model. This particular observation, however, was above the average noise level (1.8 Jy beam$^{-1}$) with an rms of $\sim$2 Jy beam$^{-1}$. Neither the transient nor ghost source had significant detections (with the significance level now reduced to 5$\sigma$ in order to try and detect the transient), and even surrounding field sources were hard to distinguish because of the poorer image quality.

In an attempt to improve the situation, using our assumption that the transient should be included in the sky model for the observation, we phase-calibrated this dataset again using a larger solution interval of 1-minute (previously 10 s), to allow more signal to noise for the calculations. Using a source extraction threshold of $5\sigma$, {\sc trap} was able to find the transient source in the second and third of the 5, 2-minute images: the flux densities are 20.9 ($8\sigma$) and 18.7 ($7\sigma$) Jy, respectively.  The light curve can be seen in Figure~\ref{fig:2minfluxes}.

We were concerned about forcing the flux of the transient to a specific value by simply entering that value into the sky model, especially as in this case the fluxes returned were approximately equal to the flux which was entered (20 Jy). Thus, we repeated this test, but this time entering a 10-Jy transient at the position. In the 10-Jy case the transient was detected in the second image only at a lower flux density. The forced fit performed by the {\sc trap} in the third image yields a flux measurement of 13 Jy (just below $5\sigma$), before dropping off, which mimics the characteristics of the 20 Jy sky model case.

The first 2-minute image from each test was of noticeably poorer quality than the other four, 2-minute images of the observation. As seen in Figure~\ref{fig:2minfluxes}, the forced extraction at the transient location in the first image returns the same flux density value (10 Jy) in each sky model test case. This value hints at the transient being present in this epoch as this flux level is higher than the fourth and fifth epochs, where the transient is no longer detected in both cases. However, due to the uncertainty in this image and the larger error bars associated with this measurement, we cannot state for certain that this is the case.

We attempted to split the dataset which had been calibrated directly from the NCP field sky model, as discussed in Section~\ref{sec:ncpghost}, but the calibration was not of sufficient quality to achieve useful results.

The results here therefore suggest that the transient was brightest between the second and sixth minute of the observation, a period of four minutes. However, we are unable to fully characterise the decay, or especially the rise time of the event, and hence we cannot rule out the transient being active over a longer, 10-minute time-scale.

\subsection{Testing if a source can be created by the sky model}
\label{sec:modeltests}
\begin{figure*}
\centering
\includegraphics[scale=0.39]{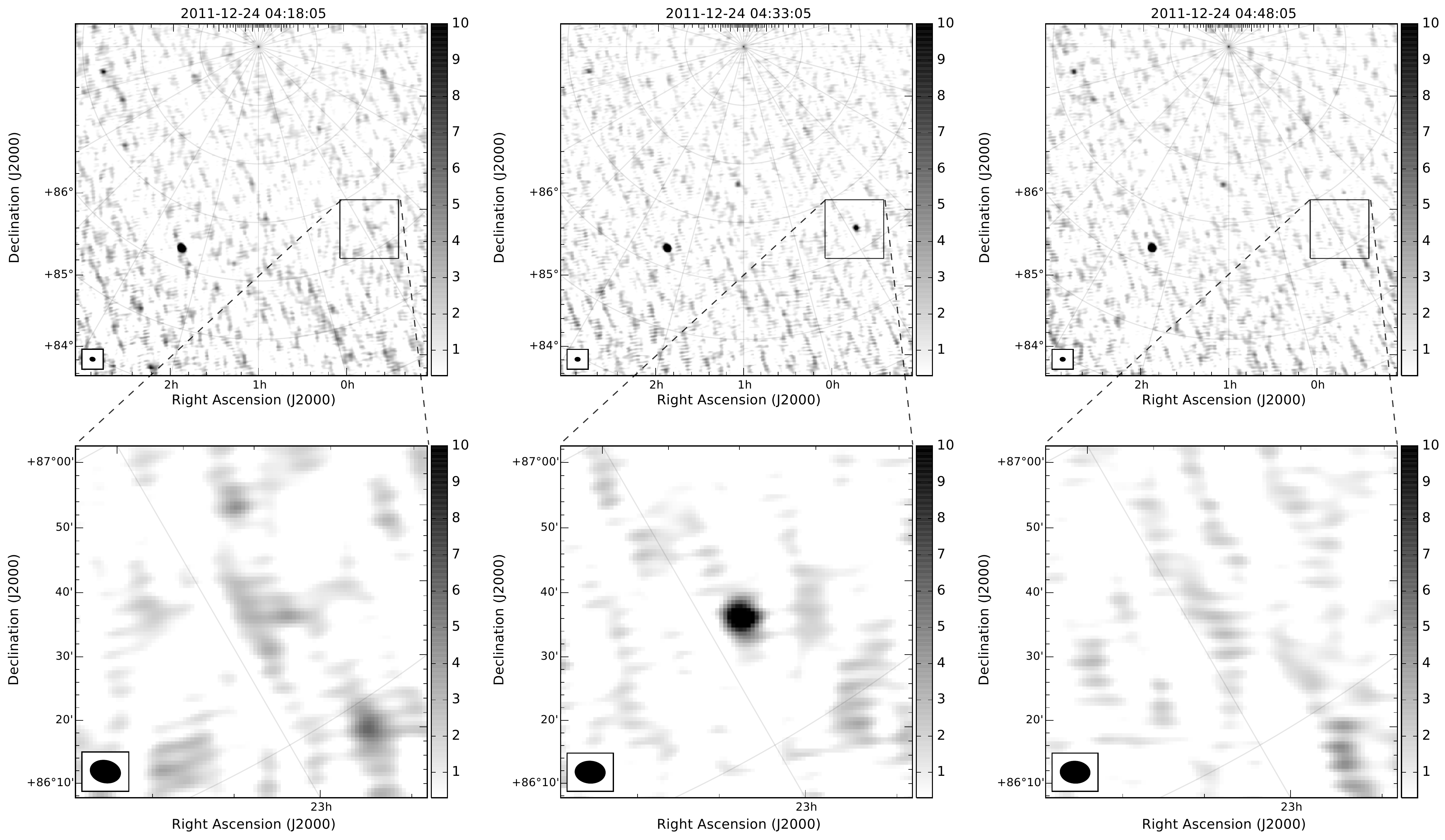}
\caption{A sequence of images in time: the transient detection image along with the snapshots before and after the event, together with a zoom-in of the transient location. Importantly, each observation has been processed with the transient included in the calibration sky model, showing how even with this taken into consideration, there are no significant detections before or after the transient. These images were created using the standard imaging parameters discussed throughout the paper, including projected baselines of up to 10 km in length. The synthesized beam can be seen in the bottom-left of each image. The colour bar units are Jy beam$^{-1}$.}\label{fig:ncpothers}
\end{figure*}

Because the transient did not correspond to any source contained in the sky model, a major concern was the possibility of `creating' false sources in the field by purely inserting them into the sky model. This could explain the apparent responsiveness of the candidate to an entry in the sky model, and perhaps a source placed anywhere in the field would have the same effect, both in creating a source and causing the ghost source to disappear. We tested this in two ways. Firstly, the snapshot containing the candidate was reprocessed with the candidate component of the sky model moved to an empty, unrelated location on the sky. This resulted in no source being `created' at this location and also left the candidate, and ghost, unaffected from their original detection states.\\
\\
The second test was to process the two preceding and two subsequent snapshots with the candidate component inserted into the sky model at its correct location. Previously, no detection was made of the candidate in any other snapshot, and as the data were recorded in sequence, the \textit{uv} coverage of these observations were all very similar. The result was that, once more, no source was present at the candidate location, even when placed in the sky model; this can be seen in Figure~\ref{fig:ncpothers}, which shows the detection of the candidate along with the snapshots before and after in time.

These two results meant that simply entering sources into the sky model at an arbitrary position would not `create' an artificial source. In contrast, the responsiveness of the transient candidate to such input at the correct position suggested it was a real source present in the data.

\subsection{Further validity testing}
\label{sec:tests}
A final set of tests and checks were performed to investigate whether ILT J225347+862146 was an unexpected artefact. With LOFAR being commissioned at the time, an artefact would not be completely surprising. While the telescope was in a good working state, a lack of optimisation of aspects such as station calibration and beam models could cause issues. A series of tests were devised to rule out certain possible artefact causes, all performed with the source both in and out of the sky model when processing. These tests were:\\

\begin{itemize}
\item \textbf{Broadband RFI} - Care was taken to manually reduce the data, removing anything left over that was suspected of being RFI, as well as running another pass of {\sc aoflagger} on the data after calibration. Neither method affected the transient source.\\
\item \textbf{Narrow-band RFI} - To rule out the possibility of narrow-band RFI, the already limited bandwidth was split into two and processed separately. The transient source remained in each half of the bandwidth, with a flux consistent within the $1\sigma$ error bars between the two halves.\\
\item \textbf{Calibrator Issues} - The calibrator observation contains nothing peculiar and was of good quality. The calibrator for this observation was 3C 295.\\
\item \textbf{Calibrator Gains Only} - Previously discussed in Section~\ref{sec:ncpghost}, this test meant the phase-only calibration step using the target field sky model was skipped; instead we imaged the dataset with the gain amplitude and phase solutions obtained directly from the calibrator being applied. ILT J225347+862146 was still present in the resulting image along with the ghost.\\
\item \textbf{Phase Centre Shift} - The phase centre of the observation was shifted to that of the transient (a shift of $\sim$4 deg). The transient was still clearly visible with the shift, especially when the source was included in the sky model.\\
\item \textbf{Equal Local Sidereal Time Observations} - Sixteen observations were found to have very similar local sidereal times (LST) to that of the detection measurement, so these were used to check whether the candidate was potentially caused by that particular projection of the baselines on the sky. There was no detection in any of these observations, which covered four months of recording.\\
\item \textbf{Bad Station Removal} - Along with the automatic tool that was part of the initial four tests, a manual inspection of the data was also carried out, which was in agreement with the results from the tool: the same two stations were perceived as bad. After these stations were flagged, the image was generally cleaner from artefacts with the transient source unaffected.\\
\item \textbf{Random Subset of Stations} - Half of the 33 stations used in the observation were randomly removed, after calibration, with the remaining data being re-imaged. This was repeated three times and the transient source continued to be present in each of three resulting maps.\\
\item \textbf{Dirty Map Check} - The source is present in the dirty map.\\
\item \textbf{Field Subtraction} - Using a sky model derived from the deeper image, the entire field apart from the transient was subtracted from the dataset. The transient and ghost were clearly visible in this case, with the same flux density.\\
\item \textbf{Imaging at a Different Resolution} - Reducing the maximum baseline length used when imaging from 10 km to various lower values had no impact on the transient. An image using a maximum projected baseline length of 15 km was also produced in which the source was still present. However, we did not consider any images produced with projected baselines longer than 10 km scientifically useful, due to concerns regarding the quality of calibration.\\
\item \textbf{Different Imaging Weighting Schemes \& Imager} - Checking for further side-lobe related issues, the imaging was redone using natural and uniform weighting. The source remained in the resulting maps. The imager itself was also checked by imaging the observation with the `Common Astronomy Software Applications' \citep[{\sc casa};][]{CASA} software rather than {\sc awimager} (this meant that no primary beam correction was made) and the source was still present. In this case, the candidate source was marginally brighter than the ghost: the flux densities were 5.3 and 4.5 Jy respectively.
\end{itemize}

\subsection{What is this transient?}
\label{sec:discussion}
With the candidate successfully passing the numerous exhaustive tests detailed previously, we concluded that the candidate was a real astrophysical source. Hence we proceeded to investigate its possible origin.

\subsubsection{Catalogue search and multi-wavelength follow-up}
\label{sec:followup}

\begin{figure}
\centering
\includegraphics[scale=0.34]{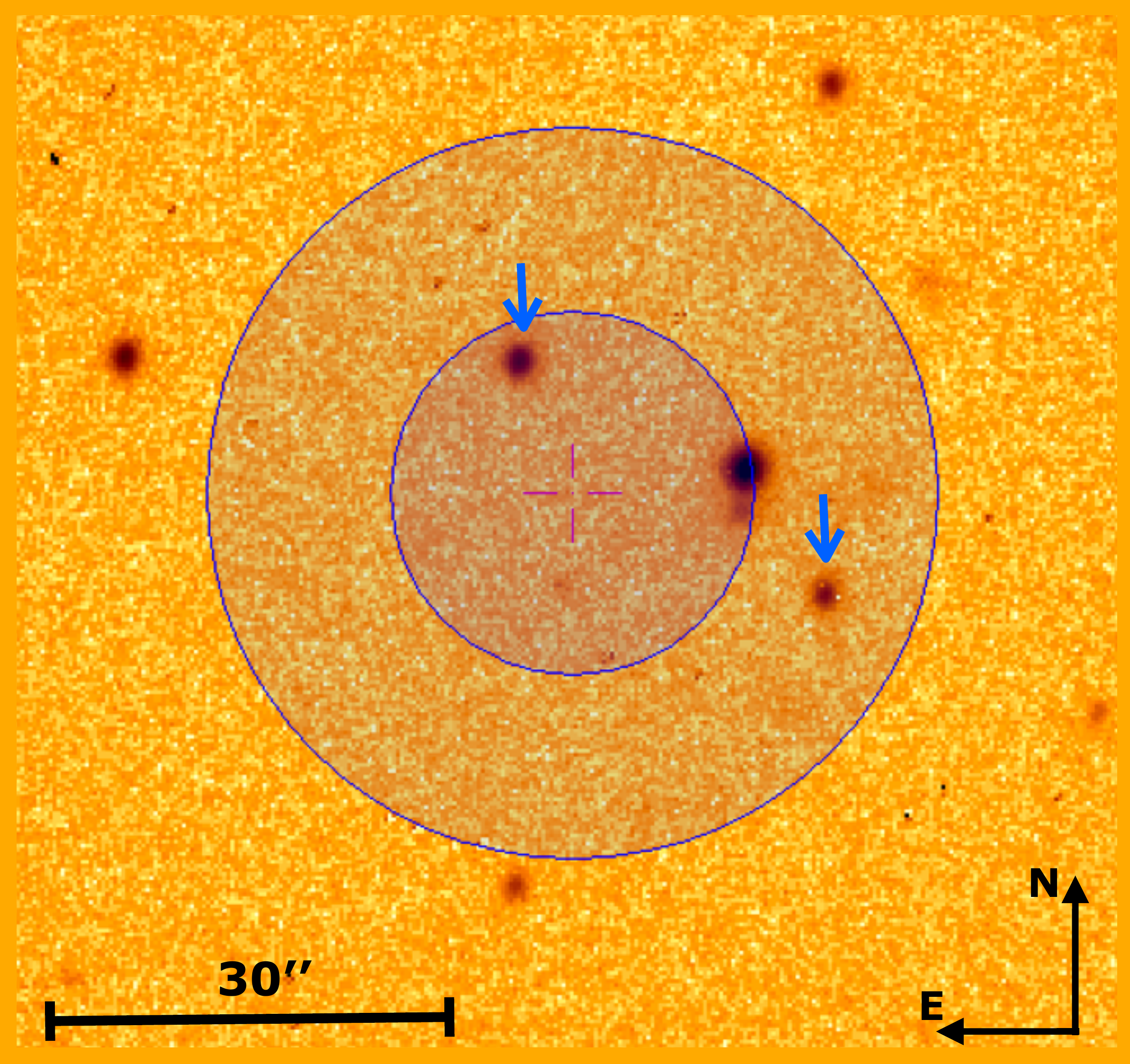}
\caption{The resultant combined optical image obtained with follow-up observations using the Liverpool Telescope. It reaches a depth of  $r' \sim$ 22--22.5 mag and is calibrated against the USNO-B1 catalogue \citep{Monet}. The inner circle marked on the image, centred at the reported LOFAR transient position, represents a $1\sigma$ positional error of radius 14 arcsec. The outer circle shows the $2\sigma$ positional error circle of radius 28 arcsec. The two stars which have proper motions higher than 100 mas year$^{-1}$, as indicated by the USNO-B1 \citep{Monet} catalogue, are indicated by arrows.} 
\label{fig:optical}
\end{figure}

No source was found within a 2-arcmin radius of the transient position in historical radio catalogues, including VLSS, WENSS and NVSS. Also, no potential counterpart or related object was found in high-energy catalogues, and no published gamma-ray burst or supernova event is known at the position.

We carried out optical follow-up of the field by using the Liverpool Telescope  \citep[LT;][]{LT}, though it is worth noting that the transient was discovered two years after the event date. Therefore, while a direct counterpart would not be observed, an object of a certain type may be identified in the vicinity of the transient, and could be potentially associated with the radio emission. These observations with the LT used the $r'$ band filter on four different dates, which are shown in Table~\ref{table:liverpool}, totalling 11 epochs and 1500 seconds of exposure time.

\begin{table}
\centering
\caption{Details of the follow-up observations performed with the Liverpool Telescope.}
\label{table:liverpool}
\begin{tabular}{|c|c|c|}
\hline
Obs. Date & \# Epochs & Epoch Exposure Time\\
 & &(s)\\
\hline
\hline
2013/11/28 & 3 & 100\\
2013/12/21 & 3 & 100\\
2013/12/30 & 3 & 100\\
2014/01/12 & 2 & 300\\
\hline
\end{tabular}
\end{table}

The combined image, shown in Figure~\ref{fig:optical}, was calibrated against USNO-B1 \citep{Monet} and reaches $r' \sim$ 22--22.5 mag. To search for a possible optical counterpart to ILT J225347+862146, we established an error on the positional measurement of the transient from the LOFAR data. This was calculated by accounting for the following uncertainties which contribute to the overall error: the error of the source extraction performed by the {\sc trap} which was reported as 11 arcsec and is measured following the error analysis of \citet{condon}; the average scatter of the extracted positions of bright sources in the dataset by {\sc trap}, measured to be 7 arcsec; and finally the reported 5 arcsec positional error of the VLSS catalogue, which the phase calibration sky model is based upon. Summing these values in quadrature we gain the final positional error of 14 arcsec.

Using this value, we unambiguously detect four sources within a $2\sigma$ error circle centred at the transient position, one of which is fully enclosed by the $1\sigma$ uncertainty as shown in Figure~\ref{fig:optical}. None of these sources displayed either strong short-term (minutes) or long-term (weeks) variability. We used the USNO-B1 catalogue to look for high-proper-motion stars in the field, with the assumption that these would be nearby objects. Two of the four sources mentioned previously within the $2\sigma$ radius, had associated proper motions of higher than 100 mas year$^{-1}$ and are marked in Figure~\ref{fig:optical}. However, neither of these sources exhibited a colour consistent with being an M-dwarf or any other possible transient object (but we note that the error associated to the USNO-B1 catalogue colours is significant). In addition, both sources were detected in our LT images and no associated variability was observed. We also consulted the Wide-Field Infrared Survey Explorer \citep[WISE;][]{wise} and the Two Micron All Sky Survey \citep[2MASS;][]{2mass}; a total of three and two sources were located within the $2\sigma$ error circle in each survey, respectively. In each case, two objects were the previously reported high-proper-motion stars; however neither of these, or the one other source in WISE, were found to be variable.

Given possible further uncertainties of the accuracy of the measured position in the LOFAR band, we extended the error circle to a radius of one arcmin from the transient position. A total of 20 sources were within this larger error circle in our LT observations; however, as previously, there was no strong evidence of a possible association with ILT J225347+862146.

We conclude that there is no obvious counterpart candidate. We now consider whether the radio emission could arise from either an incoherent or coherent process. 

\subsubsection{Incoherent origin}
\label{sec:incoherent}
If we consider the incoherent emission process, we can place a limit on the maximum distance of the source, by using the known characteristics of the transient along with assuming that its brightness temperature (T$_{\textrm{B}}$) is at the maximum T$_{\textrm{B}}=10^{12}$ K limit, T$_{\textrm{Bmax}}$, for (un-beamed) synchrotron radiation \citep{Tlim}. To do this we make use of the Rayleigh-Jeans law, that is
\begin{equation}
T_{\textrm{Bmax}} =\frac{\Delta L_{\nu}}{8 \pi k \nu^2 \Delta t^2} \textrm{ ,}
\label{eq:jeans}
\end{equation}
where $\Delta L_{\nu}$ is the change of the luminosity in time-scale $\Delta t$, $k$ is the Boltzmann constant and $\nu$ is the observing frequency. The luminosity at frequency $\nu$ is defined as $L_{\nu}=4\pi d^2S$, where $d$ is the distance and $S$ is flux density. Using this and re-arranging Equation~\ref{eq:jeans} we obtain an expression for the distance as follows:
\begin{equation}
d^2=\frac{2 k \nu^2 \Delta t^2 T_{\textrm{Bmax}}}{\Delta S} \textrm{ .}
\label{eq:distance}
\end{equation}
For the flux density change $\Delta S = 20$ Jy in time $\Delta t = 10$ min, we obtain a maximum distance of 13.7 pc. This points to the possibility of the transient being a nearby flare star.

In the case where the source is relativistic, with the synchrotron radiation now beamed, the observed brightness temperature could exceed $10^{12}$ K. This is seen in populations such as AGN \citep{agnbright1, agnbright2} and GRBs \citep{grbbright1, grbbright2}. From Equation~\ref{eq:distance} it can be seen that $d \propto T^{\frac{1}{2}}$. For example, if a brightness temperature such as $10^{16}$ K was observed, a value at which these sources can sometimes appear \citep{Gosia}, this would place the distance estimate of the transient at 1.4 kpc, an unrealistic distance for such classes of objects.\\

Various previous studies have investigated the radio transient properties of flare stars at both centimetre and decametre wavelenghts \citep{lovell,gudel,bastianreview,jackson,abdul,osten}. A recent study, \citet{boiko}, involved monitoring the flare stars AD Leonis ($d=4.9$ pc) and EV Lacertae ($d=5.1$ pc) with the UTR-2 telescope, located in Ukraine, during March 2010 and 2011. These observations, performed in the frequency range of 16.5--33 MHz, yielded a total of 167 and 73 detected radio bursts from the respective stars. In the case of AD Leonis, the average flux of the bursts was in the range of 10--50 Jy, seemingly consistent with the flux measured from ILT J225347+862146 at 60 MHz. However, one discrepancy is that the average duration of these bursts seen from AD Leonis, which is 2--12 seconds, is considerably shorter than the apparent $\sim$minutes of activity observed for ILT J225347+862146. In addition, the sole detection of the transient is possibly suspicious in this context. One would expect the detection of subsequent transient events from an active flare star over a period of four months (the time-scale of this transient search). Recent results such as \citet{superstar}, which show \textit{Kepler} solar type stars exhibiting super-flares, could offer an explanation, with some flare events being active for $\sim$10 min time-scales \citep{superflaretime}. The flare star hypothesis could be tested further by directing future observations towards the Galactic plane rather than the NCP, as the density of flare stars should be dramatically increased.

If a flare star origin is assumed for ILT J225347+862146, along with the object not not being detected in our follow-up optical image to a depth of r$' \sim 22$ mag, we can use this information to make a crude estimate of the distance of such an object such that it is consistent with a non-detection. For this, we consult a catalogue of 463 UV Cet-type flare stars compiled by \citet{flarecat}, selecting those which have a measured R band magnitude. The stars in the catalogue have a maximum distance of 50 pc, and we also only consider M type stars. These selected stars are then divided into two sub-type groups: early type stars (M0-4) and late type stars (M5-9). In total these groups have 69 and 20 stars respectively. We then calculated the average absolute magnitudes of these two groups, and subsequently at what average distance the population would be if the apparent magnitude became 22 mag. We found these distances to be $\sim2$ kpc for the early type stars and $\sim0.3$ kpc for the late type stars (consistent with estimated Galactic scale height values defined by \citealt{galacticplane}). Hence, the object responsible for ILT J225347+862146 could be further than these distances, dependent on spectral sub-type, if it is not detected in our deep optical image obtained with the LT. However, at these distances, this would make the radio event extremely luminous, which is unlikely. Therefore, it would become likely that ILT J225347+862146 is a nearby sub-stellar object. These values could be further constrained by a deeper analysis of stellar objects, so this conclusion is made tentatively. It is also worth noting that, although we were unable to pick out a possible responsible object, it is entirely possible that the source is present in the deep optical image.

\subsubsection{Coherent origin}
A new class of radio transient has emerged in recent years, known as Fast Radio Bursts (FRBs). These events are single, bright ($\sim$ Jy), bursts, which typically last for a few milliseconds and have never been seen to repeat. These bursts also exhibit high DM indicating that the population is extragalactic in origin. The first such event was seen by \citet{lorimerburst} with the Parkes Observatory in Australia. Since this discovery, a further eight bursts have been detected using Parkes \citep{keane,thornton,2014frb,RaviFRB,PetroffFRB} plus one event detected with the Arecibo Observatory \citep{arecibo}. The implied rate of FRBs based on these events has been predicted to be up to possibly thousands of FRBs occurring every day over the entire sky \citep{Hassall,lorimer2}. The progenitors of all these events are unknown, leading to a wide range of proposed theories regarding how FRBs are produced \citep[see][]{frbneutron,frbwhitedwarf,supramassive,frbgrb, frbflarestar, frbKulkarni,mottez14}. Searches at low frequencies have, thus far, not detected any FRBs, nor have any FRBs been found using interferometric arrays which would enable a better localisation of any discovered burst \citep{LOTAS,FRBVLA}.\\

\citet{Hassall} showed that when the scattering of a highly-dispersed burst is significant, imaging surveys for FRBs can be competitive with pulsar-like, high-time-resolution surveys, if not more sensitive, in detecting such events. To investigate this possibility, we consider how the scattering time and fluence of the four FRB events reported in \citet{thornton} at 1.3 GHz, compare to this event at 60 MHz. Firstly, to achieve estimates for the scattering time of the Thornton events at 60 MHz, we use the standard relation of 
\begin{equation}
\tau_{sc}(\nu) \propto \nu^{\gamma} \textrm{ ,}
\label{eq:scatt}
\end{equation}
where $\gamma=-4$. For the purposes of this scenario we ignore any dispersion effects and assume that the recorded burst duration is dominated by scattering. This assumption is quite reasonable when considering the value of any dispersion induced smearing of the signal, $\Delta t_{\textrm{D}}$, which can be calculated per MHz of bandwidth by
\begin{equation}
\Delta t_{\textrm{D}} = 8.3 \times 10^3 \textrm{ DM } \nu^{-3}_{\textrm{MHz}} \textrm{ ,}
\label{eq:dispersion}
\end{equation}
where DM is the dispersion measure and $\nu_{\textrm{MHz}}$ is the observing frequency in MHz. With a bandwidth of 183 kHz and an observing frequency of 60 MHz, $\Delta t_{\textrm{D}} = 7 \times 10^{-3} \textrm{ DM s}$. Thus, even with a DM value of 1000\,pc\,cm$^{\rm{-3}}$, $\Delta t_{\textrm{D}}$ would only cause 7 s of smearing. Table~\ref{table:frbs} shows that the predicted, scatter-dominated width of the events at 60 MHz range from 242--1234 s (taking the upper limit values), with the highest value belonging to FRB 110220. We see that the maximum duration of our transient, ILT J225347+862146, of $<$ 660 s is quite consistent with that expected from an FRB at 60 MHz. In reality, only FRB 110220 showed any evidence of scattering.

\begin{table}
\centering
\caption{The observed width of the four bursts reported in \citet{thornton} along with the estimated width of the event at 60 MHz. To calculate the estimated width we use the relation $\tau_{sc}(\nu) \propto \nu^{\gamma}$ where $\gamma=-4$. Dispersion effects are ignored and we assume a scenario of the reported widths being dominated by scattering. Of the four reported bursts, FRB 110220 was the sole event to show any evidence of scattering.}
\label{table:frbs}
\begin{tabular}{|c|c|c|}
\hline
Event & Observed width  & Estimated width\\
& at 1.3 GHz (ms)&  at 60 MHz (s)\\
\hline
\hline
FRB 110220 & 5.6 & 1234\\
FRB 110627 & $<$ 1.4 & $<$ 309\\
FRB 110703 &  $<$ 4.3 & $<$ 948\\
FRB 120127 & $<$ 1.1 & $<$ 242\\
\hline
\end{tabular}
\end{table}

Next we compare the fluence of the events. Taking a width of 11 min for ILT J225347+862146 ($6.6 \times 10^5$ ms), and the flux as 20 Jy, the fluence of ILT J225347+862146 can be stated as $20 \times 6.6 \times 10^5 = 1.3 \times 10^7$ Jy ms at 60 MHz. Taking the shorter time-scale of four minutes (as discussed in Section~\ref{sec:timesplit}) at 20 Jy gives a fluence of $4.8 \times 10^6$ Jy ms. The event with the highest fluence as reported in \citet{thornton} was FRB 110220 with 8 Jy ms at 1.3 GHz. We can compare these fluence values assuming different spectral indices. In the case of $\alpha = 0$, a direct comparison is possible, showing that the LOFAR event has a vastly greater fluence than the known FRB. Assuming $\alpha=-2$ and extrapolating the peak flux of the LOFAR event to 1.3 GHz, the fluence now becomes $2.81\times10^4$ Jy ms for the 11 min scenario. This is still much greater than FRB 110220. For the LOFAR event to be consistent with this particular burst, which is by far the highest fluence of the four reported bursts in \citet{thornton}, then a spectral index of $\alpha\sim-4.7$ would be required. This implies that the LOFAR event would be an abnormally bright FRB, even more so than the bright \citet{lorimerburst} burst at 30 Jy. The characteristic spectral index of FRBs is currently not well defined, but a value of $-4.7$ would be very steep regardless of the population. Although the time-scale of ILT J225347+862146 is consistent with a scattered FRB at low-frequencies, the required steep spectral index along with the inconsistency between the fluence of the known FRB events, casts considerable doubt regarding an FRB origin. With the exact characteristics of FRBs currently unknown, we cannot state that ILT J225347+862146 belongs to the same population.\\

It should be noted, as described by \citet{gudelcoherent}, that coherent emission from plasma processes can also occur in stellar objects such as flare stars. Such emission has also been seen from the Sun at low-frequencies; for example in type III solar radio bursts \citep{solarreview}. For this hypothesis, the arguments which were presented in Section~\ref{sec:incoherent} concerning a flare star origin also apply here. In particular, if this were the origin of ILT J225347+862146, it appears unusual to not see the event repeat over a four month period, yet we cannot rule out this possibility.

\subsubsection{Other populations}
Other populations such as AGN and X-ray binary systems were considered. However, we have insufficient evidence to confirm or rule out such classes as the origin of ILT J225347+862146.

Over the past decade, a variety of new radio transient sources have been attributed to different kinds of neutron stars. These include populations such as Rotating Radio Transients \citep[RRATs;][]{rrats} and intermittent pulsars \citep{intpulsar}, with some intermittent pulsars seen to have periods in the off-state of more than a year \citep{intpulsaryear}. It is possible that the transient reported in this paper could be an atypical isolated neutron star such as these described populations; however at this time it is not possible to present any evidence to support this hypothesis.

\section{Transient surface density \& rates}
\label{sec:limits}
\begin{table*}
\centering
\caption{Summary of the transient surface densities and general information of the results of this work (\textit{top section}) and other low-frequency ($\leq$ 330 MHz) transient surveys (\textit{bottom section}). We follow a similar approach to \citet{Ofek2011} where $\delta t$ is the time-scale of each individual epoch searched in the survey, and $\Delta t$ is the cadence time-scales(s) of the epochs observed. The value denoted by `-' signifies that we were unsure of the correct value from the literature. A time value of `cont.' means the observations were continuous. These values are those which are used in Figures~\ref{fig:2drates} and~\ref{fig:3drates}.}
\label{table:rates}
\begin{tabular}{|c|c|c|c|c|c|c|c|c|}
\hline
Survey&Telescope&$\nu$&Sensitivity&$\rho$&$\delta t$&$\Delta t$&\#&\# Detected\\
 & &(MHz)&(Jy)&(deg$^{-2}$)&&&Epochs&Transients\\
\hline
\hline
This work & LOFAR & 60 & $> 36.1 $ &$4.1 \times 10^{-7}$ & 30 s&cont.--4 months&41\,350&0\\
This work & LOFAR & 60 & $> 21.1 $ &$1.8 \times 10^{-6}$ & 2 min&cont.--4 months&9\,262&0\\
This work & LOFAR & 60 & $> 7.9 $ &$1.4 \times 10^{-5}$ & 11 min&4 min--4 months&1\,897&1\\
This work & LOFAR & 60 & $> 5.5 $ &$5.2 \times 10^{-5}$ & 55 min&4 min--4 months&328&0\\
This work & LOFAR & 60 & $> 2.5 $ &$5.3 \times 10^{-4}$ & 297 min&4 min--4 months&32&0\\
\hline
\citet{lazio} & LWDA & 74 & $> 2\,500$ &$9.5 \times 10^{-8}$ & 5 min&2 min--4 months&$\sim$1\,272&0\\
\citet{lwatrans1} & LWA1& 74 &$> 1\,440$ &$2.2 \times 10^{-9}$ \rlap{$^a$} & 5 s &cont.--1 year&$\sim$43\,056 & 2\\
\citet{Bell} &MWA& 154 & $> 5.5 $ &$7.5 \times 10^{-5}$ & 5 min&minutes--1 year&51&0\\
\citet{Dario} &LOFAR& 150 & $> 0.5$ &$10^{-3}$ & 11 min&minutes--months&151&0\\
\citet{Cendes} &LOFAR& 149 & $> 0.5$ &$10^{-2}$ & 11 min&minutes--months&26&0\\
\citet{Hym09}$^{b}$ &VLA,GMRT&235,330& $> 30 \times 10^{-3}$& 0.034 & $\sim$3 hr & days--months & - & 3\\
\citet{jaeger} &VLA&325 & $> 2.1 \times 10^{-3}$ & 0.12 & 12 hr&1 day--1 month&6&1\\
\hline
\end{tabular}
\begin{flushleft}
$^a$ Reported as $1.4 \times 10^{-2}$ yr$^{-1}$ deg$^{-2}$. Using the integration time of 5 s, this converts to $2.2 \times 10^{-9}$ deg$^{-2}$.\\
$^b$ Values for this survey are obtained from the calculations performed by \citet{asgard} which takes into account results from \citet{Hym05,Hym06,Hym09}.
\end{flushleft}
\end{table*}

No transients were found at four of the five time-scales searched, with one detection in the other. This allows us to place upper limits on the rate of low-frequency transient events on the whole sky at these time-scales. To calculate the upper limits of the surface density of transients, Poisson statistics are used, specifically:
\begin{equation}
P(n) =
e^{-\rho A}
\label{eq:rate}
\end{equation}
where $\rho$ is the surface density of sources per square degree and $A$ represents the equivalent solid angle obtained by multiplying the area of the sky surveyed, $\Omega$, by the number of epochs $N-1$. In using this approach we are also assuming isotropic probability of a transient detection, i.e. events are likely to be extragalactic in origin. We can define $P(n)=0.05$ at the Poisson 2$\sigma$ confidence level, and by rearranging Equation~\ref{eq:rate} we can obtain the respective value of $\rho$ for each time-scale, recalling that the area of sky searched in each case was 175 deg$^{2}$. For no detections $P(0)=0.05$ is used, whereas for the 11-minute time-scale this becomes $P(1)=0.05$ because of our single detection. These values can be found in Table~\ref{table:rates}.

ILT J225347+862146 provides us with $1^{+3.74}_{-0.95}$ transient events detected in the 11-minute time-scale search, using upper and lower limits at 95\% confidence as defined by \citet{erroronone}. As a 10$\sigma$ limit was used for the source extraction, the flux density limit of this search was 7.9 Jy; moreover, 1897 11-minute epochs are equivalent to 14.5 days of observations. This equates to a transient rate of $3.9^{+14.7}_{-3.7}\times10^{-4}$ day$^{-1}$ deg$^{-2}$.

However, it should be noted that the flux density limit of this rate is defined with the assumption that sources are `non-ghosted'. As seen in this work, transients with an associated ghost can be reduced in brightness when not accounted for in processing. While we are yet to exactly constrain the magnitude of the effect, we can make an estimate by taking ILT J225347+862146 as an example. In this case the source was originally detected as 7.5 Jy with an accompanying ghost, and when accounted for in processing, the minimum flux estimate was 15 Jy in the 11 minute image (see Section~\ref{sec:flux}). Therefore a minimum reduction in flux density of 50 per cent is possible, meaning that the flux density limit of the quoted rate would rise to 8 Jy for ghosted sources. For the purposes of comparing rates, in the remainder of the paper we use the ideal, non-ghosted flux density limit of 7.9 Jy.

\begin{figure}
\centering
\includegraphics[scale=0.46]{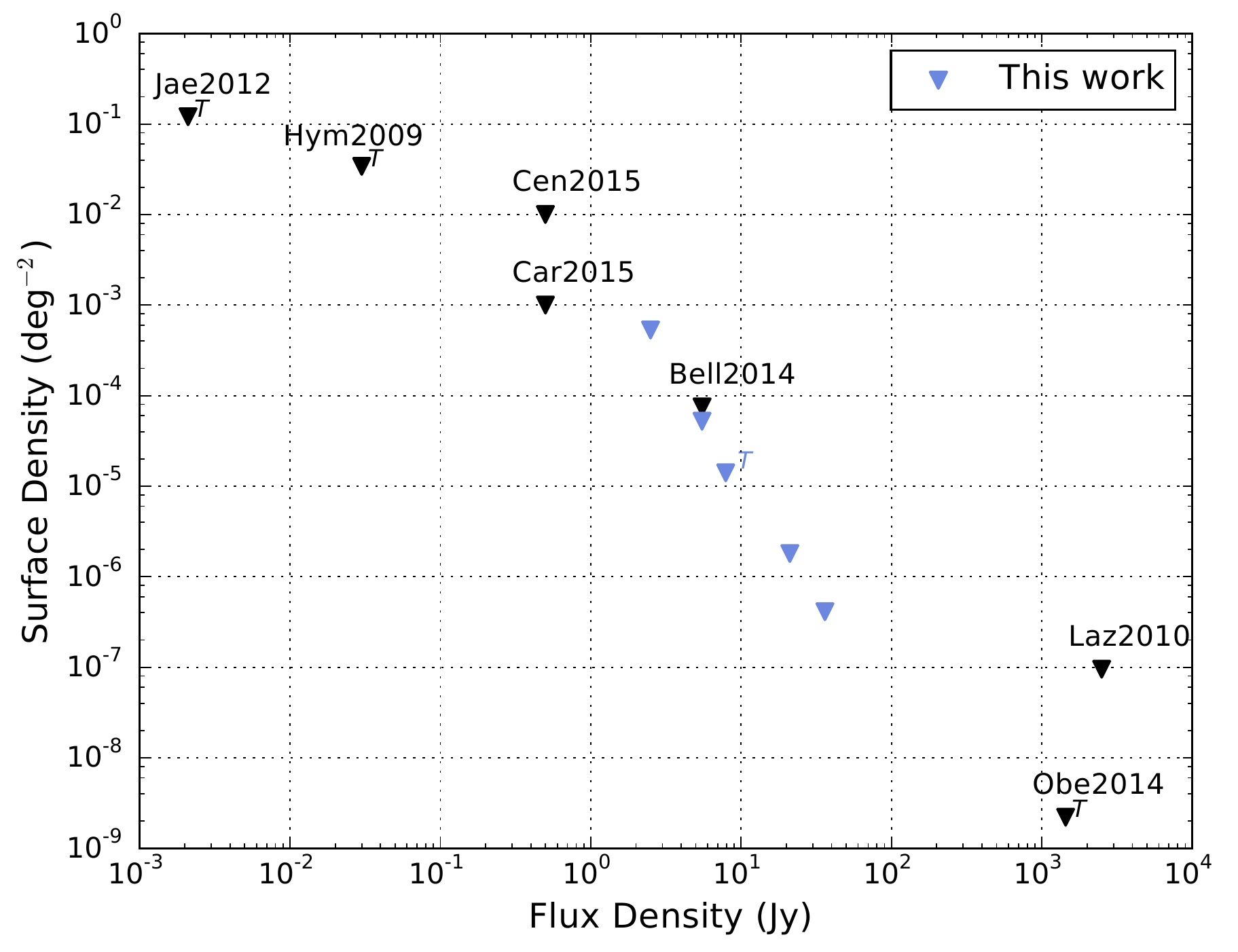}
\caption{The logarithm of the surface density (deg$^{-2}$) against the logarithm of the flux density (Jy) of low-frequency transient surveys. We do not consider variable source limits. The surface densities for which transients have been detected are marked with a \textit{T}. The surveys included are as follows: \citet{Hym05,Hym06,Hym09} (Hym2009); \citet{lazio} (Laz2010); \citet{jaeger} (Jae2012); \citet{Bell} (Bel2014); \citet{lwatrans1} (Obe2014); \citet{Dario} (Car2015) and \citet{Cendes} (Cen2015).} 
\label{fig:2drates}
\end{figure}

\begin{figure*}
\centering
\includegraphics[scale=0.34]{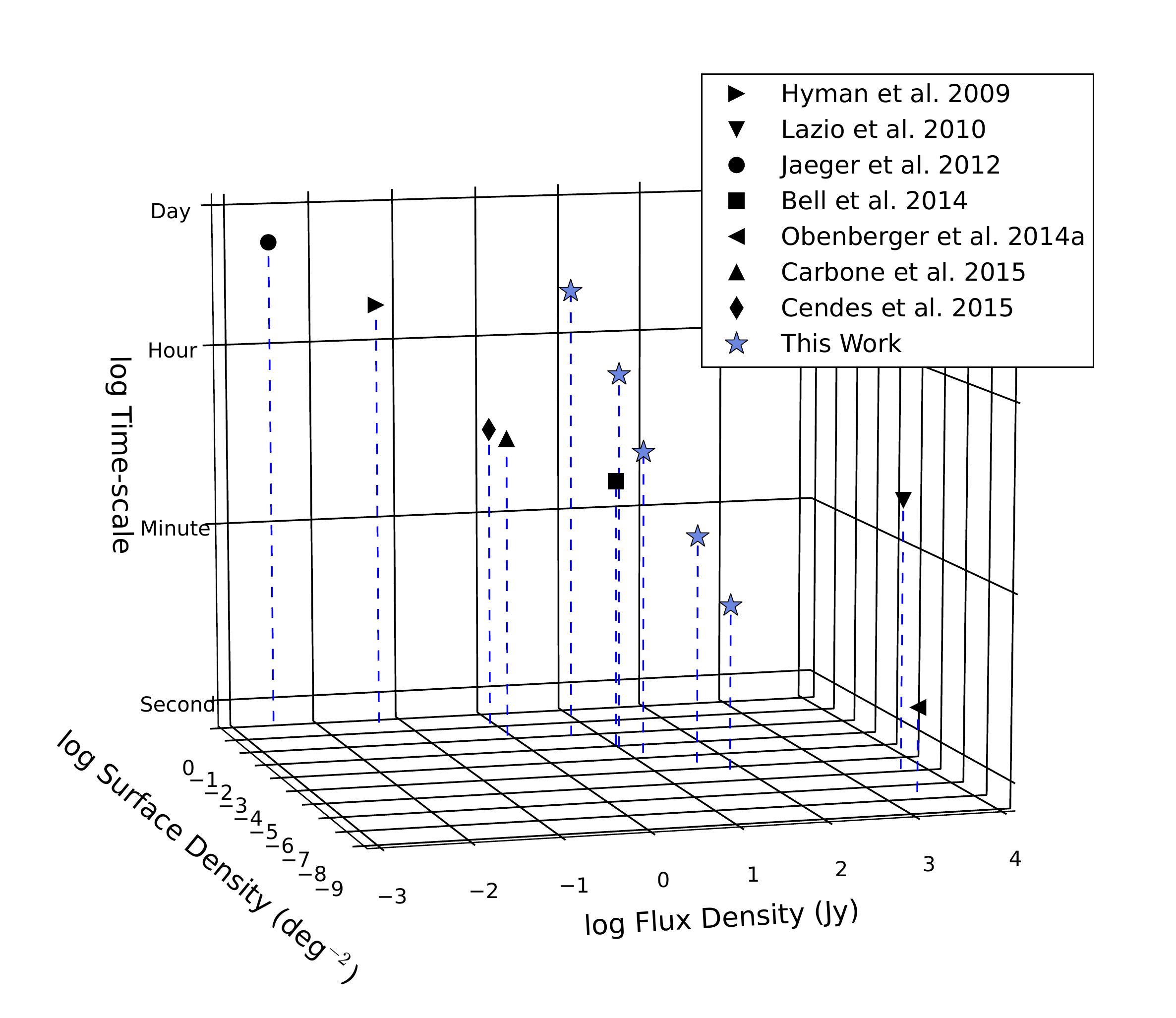}
\includegraphics[scale=0.47]{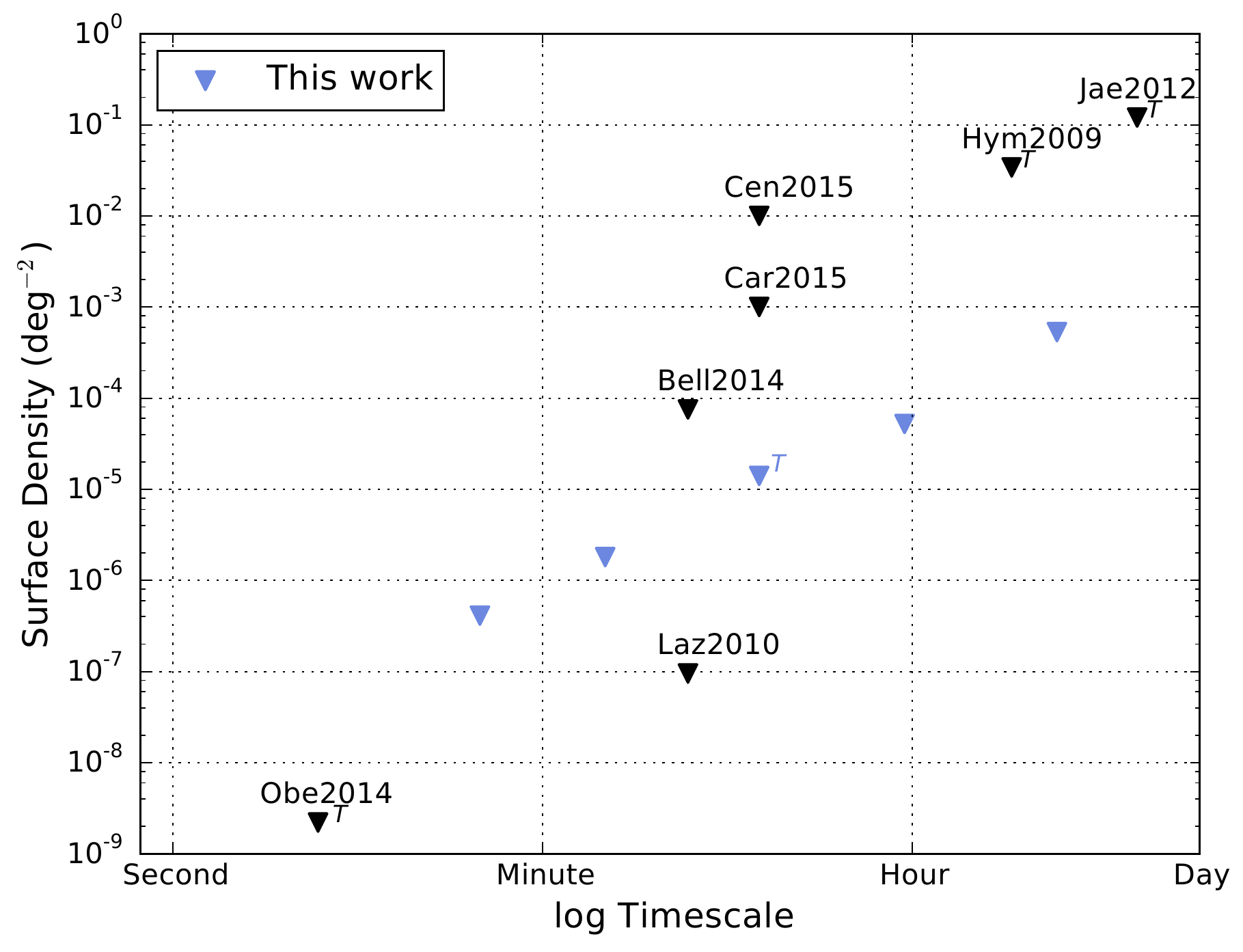}
\caption{\textit{Left panel}: A 3D plot presenting the 3D low-frequency transient search phase space when considering the time-scale of the search. Here, the logarithm of the surface density (deg$^{-2}$) is plotted against the logarithms of the flux density (Jy) and survey time-scale, with the latter on the z-axis. The viewing angle has been selected to primarily show the flux density and time-scale comparison. We do not consider variable source limits in this plot. The 11-minute limit from this work, as well as the limits of \citet{Hym09}, \citet{jaeger} and \citet{lwatrans1}, have been derived from the detection of one or more transients. The others are limits placed with no detections. \textit{Right panel}: The same plot as the left panel, but now the flux density axis has been collapsed to clearly show the surface density and time-scale comparison. The surveys included are as follows: \citet{Hym05,Hym06,Hym09} (Hym2009); \citet{lazio} (Laz2010); \citet{jaeger} (Jae2012); \citet{Bell} (Bel2014); \citet{lwatrans1} (Obe2014); \citet{Dario} (Car2015) and \citet{Cendes} (Cen2015).}\label{fig:3drates}
\end{figure*}

\subsection{Comparison to other transient surveys}
\label{sec:compare}
We primarily consider how our results relate to other low-frequency surveys ($\leq$ 330 MHz). We also only compare against transient surveys. Figure~\ref{fig:2drates} shows the results in which we have included all the surveys that are summarised in Table~\ref{table:rates}. In general, the results are consistent with the previous low-frequency surveys. We are able to improve upon the sensitivity of the \citet{lazio} surface density by at least two orders of magnitude, with the data points from this work providing some of the most extensive searches thus far at low frequencies.

It is also possible to extrapolate our results to gigahertz frequencies, for which extensive reviews of high-frequency surveys have been compiled by \citet{FenderBell}, \citet{Ofek2011} and \citet{Fender}. We find that our limits become competitive with previous surveys for transient populations with a spectral index of $-1$, and probe deeper than previous surveys if the spectral index is steepened to $-2$.

\subsection{Comparison of time-dependent surface densities}
It is important to realise that when comparing the transient surface density and flux densities of different surveys, the time-scale at which the survey was performed is just as important. A survey only looking at month-scale epochs will not be sensitive to minute or sub-minute-scale transients. The converse is also true depending on the survey length and sensitivity. Defining the sensitive time-scale of a transient survey is a complex task, with the epoch time (i.e. the time-scale of an individual epoch) and cadence of the epochs usually incorporating a range of values. Generally, we assume that the integration time of each observation is the time-scale of a transient on which the survey is most sensitive to. If the integration time matches the duration of the transient event, then the signal-to-noise will, in the majority of cases, be maximised. However as surveys are designed differently, the integration time is not always the equivalent `epoch time'. Some epochs are created by averaging many different observations together for example, or other methods such as creating mosaic fields. This can be especially true in gigahertz surveys. However, the low-frequency searches presented here as a comparison mostly do have an equal integration and epoch time. This is likely due to the large FoV of some of the facilities, minimising the need to use multiple observations to cover a large fraction of the sky. 

The cadence of the observations can also be just as valid as a defining characteristic. Taking this work as an example, the NCP search is also sensitive to slow transients that could evolve on time-scales of days, up to the maximum time between two observations of four months. Thus, to avoid the complex visual that would be required to represent all this information, we compare the surveys based upon their respective epoch times (see \citealt{Dario} for the alternate cadence comparison). The left panel of Figure~\ref{fig:3drates} shows the same plot as in Figure~\ref{fig:2drates}; however, a z-axis of time-scales has now been included to display which areas of the time-scale parameter space have been surveyed. The surveys included are the same which have previously been used and are summarised in Table~\ref{table:rates}. The right panel of Figure~\ref{fig:3drates} presents the same information but with the flux density axis collapsed, to clearly show the surface density against time-scale comparison.\\

What we see from this comparison is a clear definition of surface densities at the time-scales of minutes to hours within a range of sensitivities from the millijansky level to tens of jansky, which does improve upon the \citet{lazio} surface density. It also becomes apparent that there is a region of sensitivities -- the jansky--millijanksy regime -- that is yet to be explored at all time-scales. \citet{jaeger} is currently the only survey to have probed to a $\sim1$ mJy depth at low frequencies, and with a detected transient, this perhaps hints at the potential of further discoveries at these depths. Improving the sensitivity at shorter time-scales is also an area that could prove fruitful in transient searches.

\section{Conclusions}
\label{sec:conclusion}
In this paper, we have presented the results of a search for transient or variable sources at 60 MHz using the International LOFAR Telescope. The search was centred at the North Celestial Pole, covering 175 deg$^2$ of sky with a bandwidth of 195 kHz and conducted over the period 2011 December--2012 April. The search for transients and variables was performed using the automated, newly developed, Transients Pipeline ({\sc trap}). No transient or variable sources were discovered at time-scales of 30 seconds, 2 minutes, 55 minutes and 297 minutes. However, several candidates were discovered at the 11-minute time-scale. After extensive testing to check if these objects were due to calibration or imaging errors, one of these candidates is considered to be a real astrophysical event, based on the available data. The transient, ILT J225347+862146, was seen only in one 11 minute epoch out of 1897, implying a transient rate of $3.9^{+14.7}_{-3.7}\times10^{-4}$ day$^{-1}$ deg$^{-2}$. While complicated by the processing strategy, the flux density of the event is believed to be in the range of 15--25 Jy and was most active for an estimated time of four minutes. However, the rise or decay time-scale of the event is not sufficiently well defined such that the full duration of activity could extend to a 10-minute time-scale.\\

At present, we are unable to determine the astrophysical origin of ILT J225347+862146. There are no recorded objects at the transient position in previous radio or high-energy catalogues. Optical follow-up observations were performed at the transient position, with 20 objects detected within the 1-arcminute-radius error circle. None of these optical sources showed any short or long-term variability and no immediately obvious counterpart was identified in the field. However, the discovery of this transient two years after it was active diminishes the effectiveness of follow-up observations, highlighting the need for real-time transient searches and follow-up. 

We considered the possibility of the transient being a flare star event, due to the likely close proximity of the object. However, the time-scale of the burst is an order of magnitude longer than what would be expected from previous observations of flare stars at low frequencies. We also considered whether the event could be an FRB. While the duration of the event is consistent with a scattered burst at 1.3 GHz extrapolated to 60 MHz, it was considered unlikely due to the transient exhibiting a much larger fluence than would be expected from previously seen FRBs, which would require a very steep spectral index ($\alpha < -4.7$) to be plausible.

With LOFAR and other instruments now fully operational, the low-frequency transient sky is being probed to depths that have never previously been achieved. If the discovered transient presented in this paper is a member of a real population, then there is no question that more will be found in future and current dedicated transient surveys. This is especially true with ever improving calibration techniques and more accurate sky models at these low frequencies, combined with these surveys taking advantage of the full capabilities of these new, current generation telescopes.

\section*{Acknowledgements}
A.J.S., J.W.B., T.E.H., T.M.-D., T.D.S. and M.P. acknowledge support from the European Research Council via Advanced Investigator Grant no. 267697 4 $\pi$ sky: Extreme Astrophysics with Revolutionary Radio Telescopes (PI: R.P. Fender). T.M.-D. also acknowledges support by the Spanish Ministerio de Economia y competitividad (MINECO) under grant AYA2013-42627. A.R., J.D.S., G.J.M., D.C., Y.C. and A.J.vdH. acknowledge support from the European Research Council via Advanced Investigator Grant no. 247295 (PI: R.A.M.J. Wijers). B.S. acknowlegdes partial funding from the research programme of the Netherlands eScience Center (www.nlesc.nl). T.L.G. acknowledges support from the South African Research Chairs Initiative of the Department of Science and Technology and National Research Foundation. S.C. acknowledges financial support from the UnivEarthS Labex program of Sorbonne Paris Cit\'e (ANR-10-LABX-0023 and ANR-11-IDEX-0005-02). J.W.T.H. acknowledges funding from an NWO Vidi fellowship and from the European Research Council under the European Union's Seventh Framework Programme (FP/2007-2013) / ERC Starting Grant agreement nr. 337062 ("DRAGNET").

We would like to warmly thank LOFAR Science Support for their efforts during commissioning and the continued support of handling and processing LOFAR data. The authors would also like to thank Rachel Osten for helpful discussions during the project, along with the anonymous referee for their insightful comments and suggestions that improved the presentation of this paper.

LOFAR, the Low Frequency Array designed and constructed by ASTRON, has facilities in several countries, that are owned by various parties (each with their own funding sources), and that are collectively operated by the International LOFAR Telescope (ILT) foundation under a joint scientific policy. 

The Liverpool Telescope is operated on the island of La Palma by Liverpool John Moores University in the Spanish Observatorio del Roque de los Muchachos of the Instituto de Astrofisica de Canarias with financial support from the UK Science and Technology Facilities Council.

This research has made use of the SIMBAD database and the VizieR catalogue access tool, operated at CDS, Strasbourg, France.




\bibliographystyle{mnras}
\bibliography{ncp} 


\textit{\newline$^{1}$ Astrophysics, Department of Physics, University of Oxford, Keble Road, Oxford OX1 3RH, UK\\
$^{2}$ Physics and Astronomy, University of Southampton, Highfield, Southampton, SO17 1BJ, UK\\
$^{3}$ ASTRON, the Netherlands Institute for Radio Astronomy, Postbus 2, 7990 AA Dwingeloo, The Netherlands\\
$^{4}$ Instituto de Astrof\'isica de Canarias, 38200 La Laguna, Tenerife, Spain\\
$^{5}$ Departamento de astrof\'isica, Univ. de La Laguna, E-38206 La Laguna, Tenerife, Spain\\
$^{6}$ Anton Pannekoek Institute for Astronomy, Science Park 904, 1098 XH Amsterdam, The Netherlands\\
$^{7}$ Department of Astrophysical Sciences, Princeton University, Princeton, NJ 08544, USA\\
$^{8}$ Department of Physics \& Electronics, Rhodes University, PO Box 94, Grahamstown, 6140 South Africa\\
$^{9}$ Centrum Wiskunde \& Informatica, Science Park 123, 1098 XG Amsterdam, The Netherlands\\
$^{10}$ SKA South Africa, 3rd Floor, The Park, Park Road, Pinelands, Cape Town 7405, South Africa\\
$^{11}$ Kapteyn Astronomical Institute, P.O. Box 800, 9700 AV Groningen, The Netherlands\\
$^{12}$ CSIRO Astronomy and Space Science, PO Box 76, Epping, NSW 1710, Australia\\
$^{13}$ ARC Centre of Excellence for All-sky Astrophysics (CAASTRO), The University of Sydney, NSW 2006, Australia\\
$^{14}$ Universit\"at Hamburg, Hamburger Sternwarte, Gojenbergsweg 112, 21029, Hamburg, Germany\\
$^{15}$ Jodrell Bank Center for Astrophysics, School of Physics and Astronomy, The University of Manchester, Manchester M13 9PL,UK\\
$^{16}$ Laboratoire AIM (CEA/IRFU - CNRS/INSU - Universit\'e Paris Diderot), CEA DSM/IRFU/SAp, F-91191 Gif-sur-Yvette, France\\
$^{17}$ Station de Radioastronomie de Nan\c{c}ay, Observatoire de Paris, PSL Research University, CNRS, Univ. Orl\'eans, OSUC, 18330 Nan\c{c}ay, France\\
$^{18}$ Th\"{u}ringer Landessternwarte, Sternwarte 5, D-07778 Tautenburg, Germany\\
$^{19}$ Department of Astrophysics/IMAPP, Radboud University Nijmegen, P.O. Box 9010, 6500 GL Nijmegen, The Netherlands\\
$^{20}$ Laboratoire Lagrange, UMR 7293, Universit\'e de Nice Sophia--Antipolis, CNRS, Observatoire de la C\^ote d’Azur, 06300 Nice, France\\
$^{21}$ LPC2E - Universite d'Orleans/CNRS\\
$^{22}$ School of Physics, Astronomy and Mathematics, University of Hertfordshire, College Lane, Hatfield AL10 9AB, UK\\
$^{23}$ Max-Planck-Institut f\"{u}r Radioastronomie, Auf dem H\"ugel 69, 53121 Bonn, Germany\\
$^{24}$ SRON Netherlands Insitute for Space Research, PO Box 800, 9700 AV Groningen, The Netherlands\\
$^{25}$ Astro Space Center of the Lebedev Physical Institute, Profsoyuznaya str. 84/32, Moscow 117997, Russia\\
$^{26}$ NAOJ Chile Observatory, National Astronomical Observatory of Japan, 2-21-1 Osawa, Mitaka, Tokyo 181-8588, Japan\\
$^{27}$ Department of Astronomy and Radio Astronomy Lab, University of California, Berkeley, CA, USA\\
$^{28}$ International Centre for Radio Astronomy Research - Curtin University, GPO Box U1987, Perth, WA 6845, Australia\\
$^{29}$ Centre de Recherche Astrophysique de Lyon, Observatoire de Lyon, 9 av Charles Andr ́e, 69561 Saint Genis Laval Cedex, France\\
$^{30}$ School of Chemical \& Physical Sciences, Victoria University of Wellington, PO Box 600, Wellington 6140, New Zealand\\
$^{31}$ School of Physics and Astronomy, Monash University, PO Box 27, Clayton, Victoria 3800, Australia\\
$^{32}$ Leiden Observatory, Leiden University, PO Box 9513, 2300 RA Leiden, The Netherlands\\
$^{33}$ GEPI, Observatoire de Paris, CNRS, Universit\'e Paris Diderot, 5 place Jules Janssen, 92190, Meudon, France\\
$^{34}$ Department of Physics, The George Washington University, 725 21st Street NW, Washington, DC 20052, USA\\
$^{35}$ Harvard-Smithsonian Center for Astrophysics, 60 Garden Street, Cambridge, MA 02138, USA\\
$^{36}$ LESIA, Observatoire de Paris, CNRS, UPMC, Universit\'e Paris-Diderot, 5 place Jules Janssen, 92195 Meudon, France\\
$^{37}$ Space Telescope Science Institute, 3700 San Martin Drive, Baltimore, MD 21218, USA\\
$^{38}$ Helmholtz-Zentrum Potsdam, DeutschesGeoForschungsZentrum GFZ, Department 1: Geodesy and Remote Sensing, Telegrafenberg, A17, 14473 Potsdam, Germany\\
$^{39}$ Shell Technology Center, Bangalore, India\\
$^{40}$ University of Twente, The Netherlands\\
$^{41}$ Institute for Astronomy, University of Edinburgh, Royal Observatory of Edinburgh, Blackford Hill, Edinburgh EH9 3HJ, UK\\
$^{42}$ Leibniz-Institut f\"{u}r Astrophysik Potsdam (AIP), An der Sternwarte 16, 14482 Potsdam, Germany\\
$^{43}$ Research School of Astronomy and Astrophysics, Australian National University, Mt Stromlo Obs., via Cotter Road, Weston, A.C.T. 2611, Australia\\
$^{44}$ Max Planck Institute for Astrophysics, Karl Schwarzschild Str. 1, 85741 Garching, Germany\\
$^{45}$ Onsala Space Observatory, Dept. of Earth and Space Sciences, Chalmers University of Technology, SE-43992 Onsala, Sweden\\
$^{46}$ SmarterVision BV, Oostersingel 5, 9401 JX Assen\\
$^{47}$ Astronomisches Institut der Ruhr-Universit\"{a}t Bochum, Universitaetsstrasse 150, 44780 Bochum, Germany\\
$^{48}$ Sodankyl\"{a} Geophysical Observatory, University of Oulu, T\"{a}htel\"{a}ntie 62, 99600 Sodankyl\"{a}, Finland\\
$^{49}$ STFC Rutherford Appleton Laboratory,  Harwell Science and Innovation Campus,  Didcot  OX11 0QX, UK\\
$^{50}$ Center for Information Technology (CIT), University of Groningen, The Netherlands\\
$^{51}$ Fakult\"{a}t f\"{u}r Physik, Universit\"{a}t Bielefeld, Postfach 100131, D-33501, Bielefeld, Germany}



%
%


\bsp	
\label{lastpage}
\end{document}